\documentclass{aa}
\usepackage{graphicx}
\usepackage[labelfont=bf]{subcaption}
\usepackage{txfonts}
\usepackage{mathtools}
\usepackage{hhline}
\usepackage{rotating}

\usepackage{hyperref} 
\hypersetup{colorlinks=true,citecolor=blue}

\usepackage{siunitx} 
\usepackage{color}

\newcommand{\G}{\mathcal{G}}

\newcommand{\Msun}{M_\odot}
\newcommand{\Rsun}{R_\odot}
\newcommand{\Mearth}{M_{\oplus}}
\newcommand{\Rearth}{R_{\oplus}}
\newcommand{\aio}{a_{\rm Io}}
\newcommand{\Mio}{M_{\rm Io}}
\newcommand{\Rio}{R_{\rm Io}}
\newcommand{\MJ}{M_{\rm Jup}}
\newcommand{\RJ}{R_{\rm Jup}}

\newcommand{\Rsat}{R_{\rm sat}}

\newcommand{\Ms}{M_{\star}}
\newcommand{\Rs}{R_{\star}}

\newcommand{\Mp}{M_{\rm p}}
\newcommand{\Rp}{R_{\rm p}}

\newcommand{\Os}{\Omega_{\star}}
\newcommand{\Op}{\Omega_{\rm p}}

\usepackage[normalem]{ulem}
\usepackage{color}
\definecolor{blue}{RGB}{0,0,255}
\definecolor{red}{RGB}{255,0,0}
\definecolor{green}{RGB}{0,200,0}
\definecolor{black}{RGB}{0,0,0}
\definecolor{EB}{RGB}{48, 178, 166}

\makeatletter
\@ifundefined{thelinenumber}{}{%
}
\makeatother

\begin{document} 

\title{Survival of satellites during the migration of a Hot Jupiter}

\author{Emeline Bolmont\inst{1,2} \and Edward Galantay\inst{1} \and Sergi Blanco-Cuaresma\inst{3,4} \and Apurva V. Oza\inst{5,6}  \and Christoph Mordasini\inst{7}}

\offprints{E. Bolmont,\\ email: emeline.bolmont@unige.ch}

\institute{
$^1$ Observatoire de Gen\`eve, Universit\'e de Gen\`eve, Chemin Pegasi 51, 1290, Sauverny, Switzerland \\
$^2$ Centre pour la vie dans l'Univers de l'Université de Genève, Genève, Switzerland \\
$^3$ Harvard-Smithsonian Center for Astrophysics, 60 Garden Street, Cambridge, MA 02138, USA \\
$^4$ Faculty of Psychology, UniDistance Suisse, Brig, Switzerland \\
$^5$ Division of Geological and Planetary Sciences, California Institute of Technology, Pasadena, USA \\
$^6$ Jet Propulsion Laboratory, California Institute of Technology, Pasadena, USA \\
$^7$ Physikalisches Institut, Universit\"at Bern, Gesellschaftsstr. 6, 3012, Bern, Switzerland}

  \date{Submitted to A\&A}

  \abstract
   {
   We investigate the origin and stability of extrasolar satellites orbiting close-in gas giants, focusing on whether these satellites can survive planetary migration within a protoplanetary disk.
To address this question, we used \textsc{Posidonius}, an N-Body code with an integrated tidal model, which we expanded to account for the migration of a gas giant within a disk. 
Our simulations include tidal interactions between a $1~\Msun$ star and a $1~\MJ$ planet, as well as between the planet and its satellite, while neglecting tides raised by the star on the satellite.
We adopt a standard equilibrium tide model for the satellite, planet, and star, and additionally explore the impact of dynamical tides in the convective regions of both the star and planet on satellite survival.
We systematically examine key parameters, including the initial satellite-planet distance, disk lifetime (which serves as a proxy for the planet’s final orbital distance), satellite mass, and satellite tidal dissipation.
For simulations incorporating dynamical tides in both the planet and star, we also explore three different initial stellar rotation periods.
Our primary finding is that satellite survival is rare if the satellite has nonzero tidal dissipation.
Survival is only possible for initial orbital distances of at least $0.6$ times the Jupiter-Io separation and for planets orbiting beyond $\approx$0.1~AU.
Satellites that fail to survive are either 1) tidally disrupted, as they experience orbital decay and cross the Roche limit, or 2) dynamically disrupted, where eccentricity excitation drives their periastron within the Roche limit.
Satellite survival is more likely for low tidal dissipation and higher satellite mass.
Given that satellites around close-in planets appear unlikely to survive planetary migration, our findings suggest that if such satellites do exist (as has been recently suggested), another process should be invoked.
In that context, we also briefly discuss the claim of the existence of a putative satellite around WASP-49 A b.
   }

   \keywords{Planet-star interactions -- Protoplanetary disks
-- Planet-disk interactions -- Planets and satellites: terrestrial planets -- Planets and satellites: interiors -- Planets and satellites: dynamical evolution and stability -- Methods: numerical -- WASP-49 A b}

   \maketitle

\section{Introduction}

Natural satellites are abundant in the Solar System, yet identifying their equivalents around exoplanets remains a challenge. Despite the rapid advancement of exoplanet detection techniques, finding evidence for exomoons has proven difficult. 
In recent years, several methods have been proposed, including direct detection by transit \citep{1999A&AS..134..553S} and indirect approaches that infer their presence through transit-timing variations caused by their gravitational influence on the host planet \citep{2009MNRAS.392..181K}. 
Another approach involves detecting the possible toroidal atmosphere of the host planet, which is composed of material ejected from the moon \citep[or rings,][this method is analogous to observations of the Jovian and Saturnian systems]{2006PASP..118.1136J}.
A relatively recent claim of detection was made in 2018 by \citet{2018SciA....4.1784T}, who discussed that the transit signal obtained with the Hubble Space Telescope was compatible with a Neptune-size exomoon orbiting a super Jupiter.
However, \citet{2019ApJ...877L..15K} re-analyzed the data and found that a moon-less model led to a better fit, concluding that an artifact of the data reduction was likely responsible for the exomoon transit signal reported by \citet{2018SciA....4.1784T}.
Not technically an exomoon detection, but a sign of circumplanetary disk was proposed by \citet{2021ApJ...916L...2B}, suggesting that satellite formation should occur for exoplanets. 
A more recent potential candidate was announced by \citet{2022NatAs...6..367K}, who claimed the detection of a 2.6~Earth radius satellite at 16 planetary radii from a host Jupiter-sized planet. 
However, its existence was questioned by \citet{2024NatAs...8..193H}.
One of the most recent tentative discoveries of an exomoon was proposed by \citet{2024OJAp....7E.109P}, who measured the obliquity of $\beta$ Pictoris b and propose that the origin of its high value might be due to the gravitational interaction with a massive moon \citep[similar to what was proposed for Saturn by][]{2021NatAs...5..345S}. 
{Finally, evidence of a much smaller lunar-sized satellite (assumed to have a radius of approximately Io's, i.e. exo-Io) around the planet WASP-49~A~b in transit spectroscopy was suggested by \citet{2019ApJ...885..168O}. 
It remains unclear whether such a satellite could form in-situ (similar to hot close-in gas giant planets, e.g. \citealt{Batygin2016}) at such close-in distances given the increased planetary radius during gas giant formation \citep[e.g.][]{2018haex.bookE.143M}. 
Recently, a recent Doppler redshift detection suggests this satellite may currently be on an $\sim$ 8 hour orbit \citep{2024ApJ...973L..53O}, based on ongoing high-resolution spectra measurements of its alkali mass flux.  
The existence of this putative satellite raises many questions about its formation and evolution.}
{A recent confirmation of the Doppler-shifted sodium signature of WASP-49 A b I was made by KECK/HIRES \citep{Unni2025}. Although the neutral sodium signature of the exomoon is suggestive, the orbital period remains unconstrained from 6-11 hour circular orbits \citep{2025A&A...694L...8S} encouraging more dynamical scrutiny.} 
A potential satellite around WASP-39 b has also been suggested by \citet{2025MNRAS.tmp.1555O}, using JWST observations.

{In this vein}, many studies look at the possibility of satellite survival around gaseous planets {or sub-Neptunes} \citep[e.g.][]{2020ApJ...902L..20Q,2021PASP..133i4401D,2023MNRAS.520..761H,2023A&A...672A..78M,2025MNRAS.542L.144P}.
The survival of satellites mainly depends on their tidal interaction with the host planet.
Indeed, satellites are close enough to their host planet to experience tidal forces: the tide raised by the planet on the satellite (satellite tide) and the tide raised by the satellite on the planet (planetary tide). 
The satellite tide is important for satellites on eccentric orbits, with a non synchronous rotation and an obliquity. 
Most of the time, the satellite tide leads to a decrease of the eccentricity, accompanied by an inward migration.  
The planetary tide is potentially important in wider range of configurations, and the outcome of the evolution depends on where the satellite is relative to the corotation radius: the orbital distance at which the satellite's orbital frequency matches the rotation frequency of the planet.
If the satellite is beyond this corotation radius, it will migrate outwards (like the moon or the satellites of Saturn, \citealt{2012Sci...338.1196C}).
However, if the satellite is inside the corotation radius, it will migrate inward and eventually fall onto the planet \citep[e.g. Phobos,][]{1981MNRAS.194..365M}.

The timescale of the evolution depends on the characteristics of both planet and satellite, such as their mass, radius, and their internal structure.
Depending on their internal structure, particularly whether they have solid and/or liquid layers, either the equilibrium tide, the dynamical tide, or both can significantly impact the evolution.
On one hand, the equilibrium tide is the result of the hydrostatic adjustment of an extended body perturbed by the presence of another body.
One of the most used equilibrium tide model is the constant time lag model \citep[or CTL model, e.g.][]{1979M&P....20..301M,1981A&A....99..126H,1998ApJ...499..853E,2010A&A...516A..64L,2011A&A...535A..94B,2015A&A...583A.116B,2020A&A...635A.117B}, which assumes that the bodies are weakly viscous.
It is thought to accurately describe the response of gas giants, but is less effective for rocky bodies \citep[in particular, it cannot reproduce the spin-orbit resonances, as observed for Mercury, e.g.][]{2013ApJ...764...27M}.
On the other hand, the dynamical tide is a wave-like response to the perturbing potential \citep[e.g.][]{1975A&A....41..329Z}. 
In particular, in the convective regions of stars or in the gaseous layers of giant planets, inertial waves can propagate \citep[e.g.][]{2004ApJ...610..477O,2007ApJ...661.1180O,2014A&A...566L...9G}. 
These waves contribute to additional dissipation, which can be several orders of magnitude higher than that of the equilibrium tide \citep[e.g.][]{2015A&A...580L...3M}. 
This additional dissipation can lead to a tidal evolution different from what we would expected based solely on the equilibrium tide \citep[][]{2016CeMDA.126..275B,2017A&A...604A.113B,2017A&A...604A.112G,2018A&A...619A..80G}. 
\citet{2016CeMDA.126..275B} showed that for a hot Jupiter orbiting a fast rotating star, the {stellar} dynamical tide could lead to a significant outward migration of the planet.
Similarly, a satellite around a Jupiter-size planet can excite tidal inertial waves in the fluid layers of the planet \citep[{e.g. for Jupiter,}][]{2024A&A...682A..85D}.

The dissipation in Jupiter is well-constrained thanks to accurate and long-term astrometric measurements of the satellites' orbits \citep[e.g.][]{2016CeMDA.126..145L}\footnote{The dissipation in Io is also well-characterized thanks to recent Doppler measurements by Juno (and Galileo): $Q_\mathrm{Io} = 11.4 \pm 3.6$ \citep{Park2025}}. 
In particular, \citet{2009Natur.459..957L} {measured $Q_\mathrm{Jup} \approx 10^4$ for the frequency of Io}, a low value {(compared to past estimates, see \citealt{1977Icar...30..301G} and the next paragraph)} likely driven by the dynamical tide in the planet's fluid envelope \citep{2004ApJ...610..477O}.
Using a {similar value of $Q_{\rm HJ} \approx 10^5$,} some studies proposed that satellites around hot Jupiters may not survive over Gyr timescales as they gradually spiral into their host planet \citep[e.g.][]{2002ApJ...575.1087B}.

However, depending on whether the dynamical tide is excited or not, estimates of the tidal dissipation can vary by several orders of magnitude.
Estimates of the turbulent viscosity of Jupiter led \citet{1977Icar...30..301G} to suggest a tidal dissipation factor for Jupiter to be $Q_\mathrm{Jup} \approx 10^{12}$ (compatible with a {very} weak equilibrium tide), which has also been discussed in detail by \citet{2004ApJ...610..477O, 2005ApJ...635..688W, 2007ApJ...661.1180O}. 
This value is significantly larger than {the tidal $Q$ measured of} Jupiter or Saturn, implying weak tides and negligible satellite migration. 
Because a hot Jupiter is expected to be in synchronous rotation with its orbital motion around the star, \citet{2009ApJ...704.1341C} argued that a satellite cannot excite the dynamical tide in the planet.
Using a similarly high value of $Q_{\rm HJ}$, they showed that a hot Jupiter could host a satellite as massive as Earth over Gyr timescales. 
Notably, \citet{2009ApJ...704.1341C} included the gravitational influence of the star on the satellite in their study. 
{Another example is \citet{2019ApJ...885..168O} who suggested} a tidal $Q_{\rm HJ}$ could be as high as $\sim$ 10$^{10}$ based on the lack of Na I detections at close-in orbits, and dynamical arguments from \citet{1977Icar...30..301G}. 
Hence, the choice of planetary dissipation is key to study the survival of the satellites.

\begin{figure}[h]
\centerline{\includegraphics[width=\columnwidth]{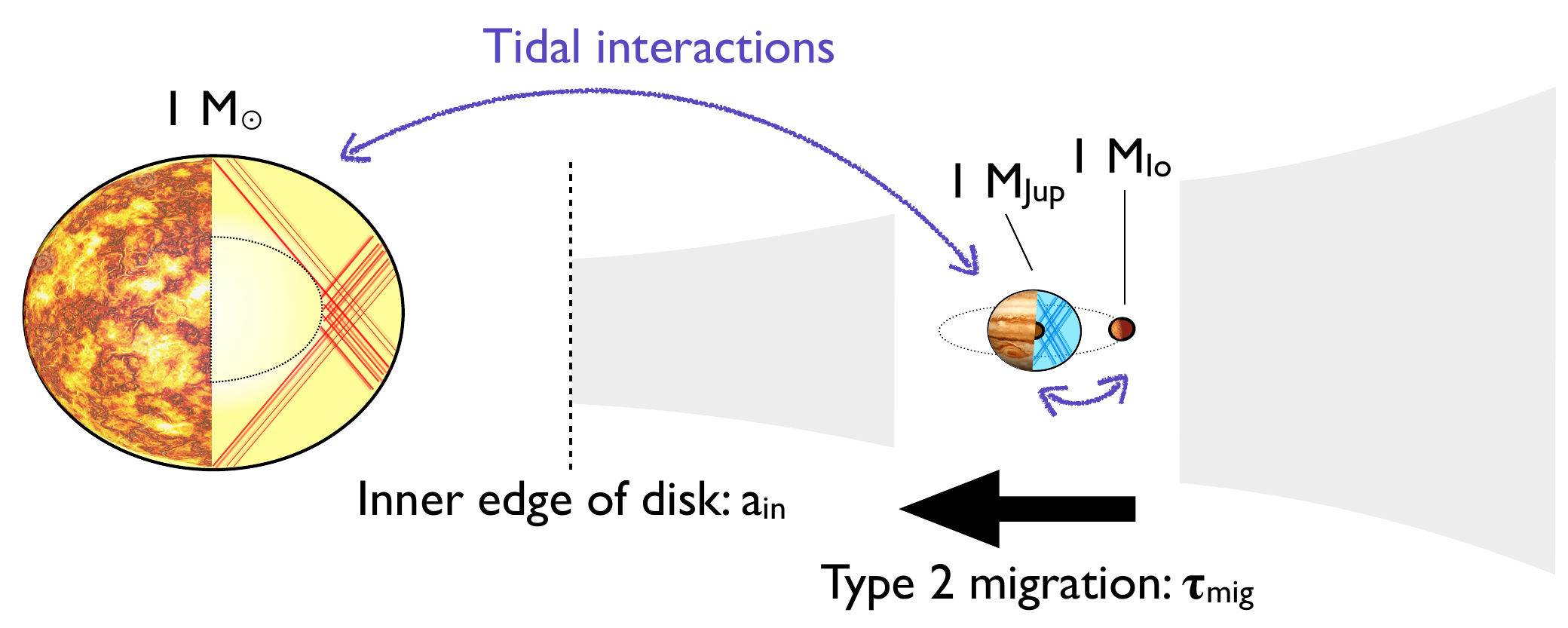}}
\caption{Schema of the simulation set-up: A Io-like satellite orbits around a Jupiter-like planet with a solar-like host star.}
\label{Fig1}
\end{figure}

We consider here the dynamical evolution of a satellite, previously formed around a Jupiter that subsequently migrates inwards to {less than} $0.1$~AU from its star (see Fig.~\ref{Fig1} for a schematic set-up). 
{While one study has examined the fate of moons around hot Jupiters in the high-eccentricity migration scenario \citep{2020MNRAS.499.4195T}, our focus here is on the fate of moons around around a Jupiter-mass planet undergoing Type-II migration.}
The simulations we present aim to assess the likelihood, based on our assumptions, of {satellite survival during the final stages of migration of the planet while accounting for the influence of the dynamical tide in both the star and planet.}

\section{The model}\label{Model}

To describe the evolution of the planet-satellite system, several mechanisms should be considered. 
First, the gravitational interactions between each body (Section~\ref{N-body}); second, the interaction between the planet and the disk, which drives the system's evolution (Section~\ref{disk_model}); and finally, the tidal interactions between the planet and the satellite, as well as between the planet and the star (Section~\ref{tidal_model}).

\subsection{N-body integration}\label{N-body}

We use the open-source N-body code \textsc{Posidonius} (\citealt{2017ewas.confE...8B,2020A&A...635A.117B}, \citealt{2021A&A...651A..23G}, \citealt{2024A&A...691L...3R} Blanco-Cuaresma \& Bolmont, in prep), which computes the evolution of multi-planet systems while accounting for various additional effects, including tidal interactions (see Section~\ref{tidal_model}) and the disk interaction (see Section~\ref{disk_model}).
 
\textsc{Posidonius} includes several evolutionary tracks for different objects such as solar-like stars (e.g. \citealt{2017A&A...604A.112G}) and for Jupiter-mass planets (\citealt{2013NatGe...6..347L}).
These tracks allow to account for the radius evolution of the star and giant planet as they age. 
As tidal interactions have a strong dependence on the radii of the bodies, it is therefore important to account for that evolution {(see Section~\ref{tidal_model})}.

\textsc{Posidonius} provides different integration schemes, and here we use IAS15 \citep{2015MNRAS.446.1424R}, which is an adaptive timestep Gauss-Radau quadrature integrator.
Unlike symplectic integrators, this integrator does not assume the existence of a dominant massive body (e.g., the host star) at a fixed reference position, such as the origin of the coordinate system. 
This flexibility allows us to configure the simulation so that tidal interactions between the star and the planet, as well as between the planet and the satellite, are both consistently computed, without requiring modifications to the existing tidal model. 
However, IAS15 is computationally expensive and significantly slower than a symplectic integrator like WHFast (\citealt{2015MNRAS.452..376R}, which is also included in \textsc{Posidonius}). As a result, the duration of our simulations is limited, with the longest simulations only extending up to 5-6 million years. 

 \subsection{Migration of the hot Jupiter} \label{disk_model}

\textsc{Posidonius} accounts for planetary migration following the planet type-II migration prescription of \citet{2018haex.bookE.143M} and \citet{2013A&A...558A.109A}.
{In this work, however, we are not looking to simulate the full migration from a few AU down to a few hundredth of AU. 
We only focus on the last part of it (from 0.15~AU on), as it is where the gravitational effect of the star starts to be important for the evolution of the satellites.
As the order of magnitude for the satellite accretion in a circumplanetary disk is of the order of $\sim10^4-10^5$~yr \citep[e.g.][]{2002AJ....124.3404C,2020A&A...633A..93R,2020ApJ...894..143B}, we consider the satellites to be fully formed by the time the planet reaches 0.15~AU.
This also allows us not to account for a circumplanetary disk, which is out of the scope of this study.}

The acceleration associated with planetary migration in the disk is given as \citep{2007A&A...472.1003F, 2013A&A...558A.109A}
\begin{equation}\label{acc_mig}
\mathbf{a}_{\rm mig} = -\frac{\mathbf{v}}{\tau_{\rm mig}}
\end{equation}
where $\mathbf{v}$ is the velocity vector of the planet, and 
$\tau_{\rm mig}$ is the migration timescale, defined as 
\begin{equation}\label{tau_mig}
\tau_{\rm mig}  = -a_p/\dot{a}_p,
\end{equation}
where $a_p$ is the semi-major axis of the planet and $\dot{a}_p$ its time derivative. 
Similarly, the accelerations associated with eccentricity damping and the inclination damping are given by
\begin{equation}\label{acc_de}
\mathbf{a}_{\rm de} = -2\frac{(\mathbf{v}.\mathbf{r})\mathbf{r}}{r^2 \tau_{\rm e}},
\end{equation}
and
\begin{equation}\label{acc_di}
\mathbf{a}_{\rm di} = -2\frac{(\mathbf{v}.\mathbf{k})\mathbf{k}}{\tau_{\rm i}},
\end{equation}
where $\mathbf{r}$ is the position vector of the planet, $r$ is its norm, $\tau_{\rm e}$ and $\tau_{\rm i}$ are the eccentricity and inclination damping timescales, and $\mathbf{k}$ is the vertical unit vector.
Here, we assume that both the eccentricity and inclination damping timescales are equal to $0.1~\tau_{\rm mig}$ \citep{2013A&A...558A.109A}. 

To compute the migration, eccentricity, and inclination damping accelerations, one needs to estimate the migration timescale $\tau_{\rm mig}$, and therefore the time derivative of the semi-major axis, $\dot{a}_p$.
{For type-II migration, we assume that the time derivative of the semi-major axis (i.e. the migration rate, $\dot{a}_p$) is limited by viscous transport within the disk \citep{2005A&A...434..343A,2018haex.bookE.143M} and is therefore given by}
\begin{equation}\label{mig_rate}
\dot{a}_p = u_r \min{\left(1,~\frac{2 \Sigma_g(r,~t) a_p^2}{M_p}\right)}
\end{equation}
where $u_r$ is the local radial velocity of the gas, $\Sigma_g(r,~t)$ is the surface density of the gas, and $M_p$ is the mass of the planet.  
The radial velocity of the gas is given by $u_r  = -3\nu/(2a_p)$,  where $\nu$ is the viscosity of the disk.
We use an $\alpha$-disk prescription {\citep{1973A&A....24..337S}} so that the $\alpha$ viscosity {parameter} is given by
\begin{equation}\label{viscosity}
\nu = \alpha c_s H = \alpha c_s^2/\Omega,
\end{equation}
where $H$ is the vertical scale height, $\Omega$ is the Keplerian frequency, and $c_s$ is the speed of sound.
{The $\alpha$ viscosity parameter is usually between $10^{-3}$ and $10^{-2}$ \citep{2009A&A...501.1161M} and in this study we choose a value of $10^{-2}$ (see Table~\ref{tab:table1}).}
The speed of sound is given by  
\begin{equation}\label{speed_of_sound}
c_s = \sqrt{\frac{k_B T}{\mu m_H}},
\end{equation}
where $k_B$ is the Boltzmann constant, $T(r)$ the temperature of the disk at $r$, $\mu$ the mean molecular weight, and $m_H$ the mass of a hydrogen atom.
The temperature of an optically thin disk is given by \citep{2004ApJ...604..388I}
\begin{equation}\label{temperature}
T(r) = 280 {\rm K} \left(\frac{r}{1 {\rm AU}}\right)^{-1/2}\left(\frac{\Ms}{\Msun}\right).
\end{equation}
Finally, we adopt the protoplanetary disk density profile from \citet{1974MNRAS.168..603L} and assume a simple exponential decrease to mimic the disk's lifetime. 
The corresponding surface disk density is given by
\begin{equation}\label{surf_density}
\Sigma_g(r,~t) = \Sigma_g(t=0,~r)~\exp\left(-t/\tau_{\rm disk}\right),
\end{equation}
where $r$ is the radial distance to the star, $t$ is the time, $\tau_{\rm disk}$ is the disk lifetime, $\Sigma_g(t=0,~r)$ is the initial density profile given by
\begin{equation}\label{initial_surf_density}
\Sigma_g(t=0,~r) = \Sigma_0 \left(\frac{r}{1{\rm AU}}\right)^{-1}~\exp\left[-\frac{r}{R_{\rm out}}\right]\left(1-\sqrt{\frac{r}{R_{\rm in}}},\right)
\end{equation}
where $\Sigma_0$ is a reference disk density, and $R_{\rm in}$ and $R_{\rm out}$ are respectively the inner radius and the characteristic outer radius of the disk (the disk does not stop at $R_{\rm out}$). 
The default parameters are listed in Table~\ref{tab:table1}.

\begin{table}[h!]
  \begin{center}
    \caption{Disk parameters.
    } 
    \label{tab:table1}
    \begin{tabular}{l|c|r} 
      Parameter	 	& Value(s) 				& Units \\
      \hline
      \hline
      $R_{\rm in}$ 	& 0.01 					& AU \\
      $R_{\rm out}$ 	& 100 					& AU \\
      $\Sigma_0$ 	& 1000 	& g/cm$^2$ \\
      $\tau_{\rm disk}$	& 5000,...,50000 			& yr\\
      $\mu$ 		& 2.4 					& - \\
      $\alpha$		& ${10^{-2}}$						& - \\
    \end{tabular}
    \tablefoot{The disk lifetime does not have a default value, it is a proxy for the distance at which the planet stops its evolution.}
  \end{center}
\end{table}

\begin{figure}[h]
\centerline{\includegraphics[width=\columnwidth]{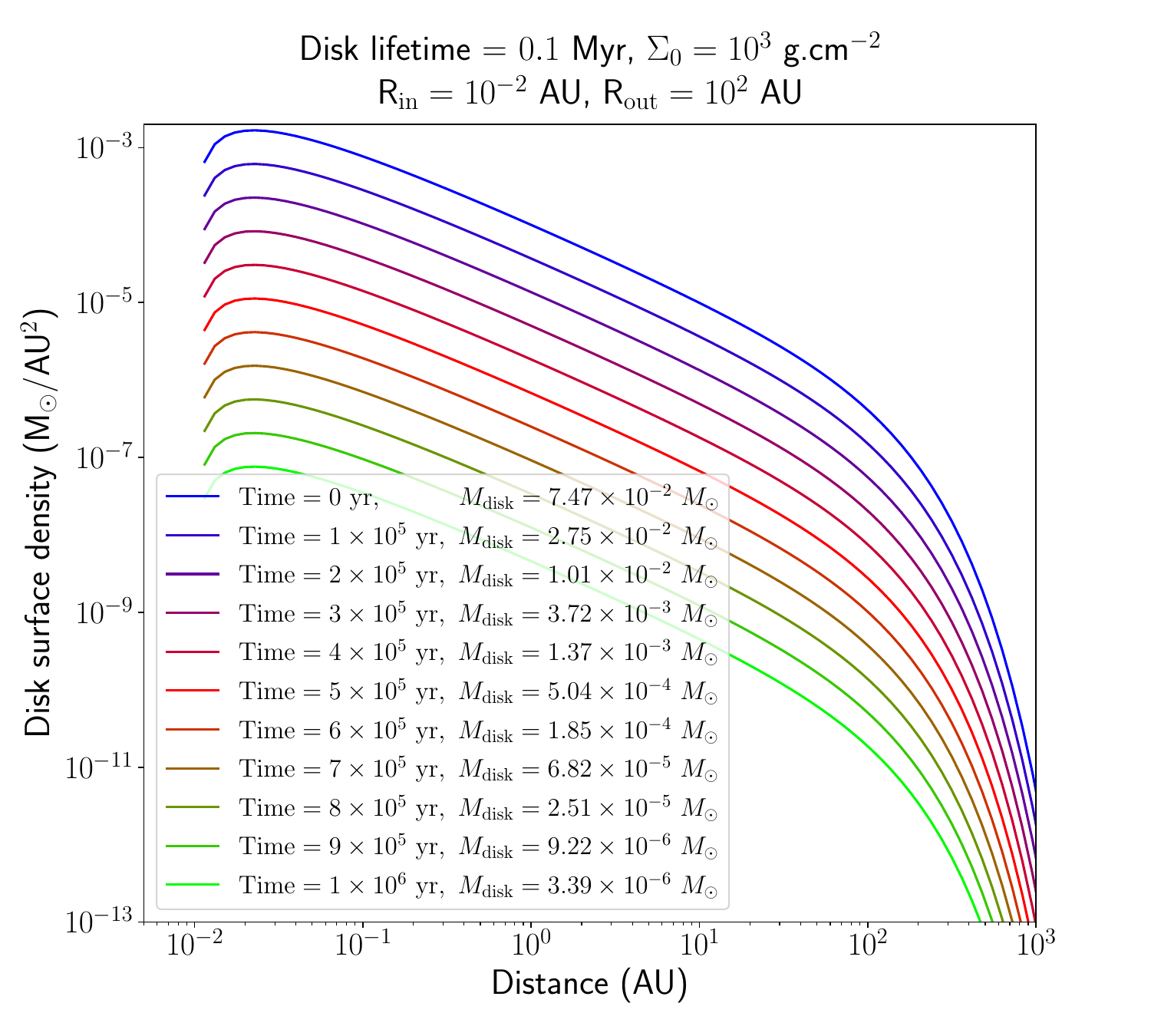}}
\caption{Evolution of the disk surface density $\Sigma_g(r,~t)$.}
\label{Fig2}
\end{figure}

The evolution of the density profile over time is illustrated in Figure~\ref{Fig2} for a disk of $\tau_{\rm disk} = 0.1$~Myr, $R_{\rm in} = 0.01$~AU and $R_{\rm out} = 100~$AU, and $\Sigma_0 = 1000~$g/cm$^2$.
Initially, the profile exhibits a maximum close to the inner edge, followed by a gradual decrease towards the outer parts of the disk. 
For radii higher than $R_{\rm out}$, the decrease becomes significantly steeper.
The density, and consequently the mass, of the disk decrease with time according to Eq~\ref{surf_density}. 
{Note that in the framework that we are using for mimicking a type-II migration \citep{2005A&A...434..343A,2018haex.bookE.143M}, there is no gap in the surface density profile.}

\begin{table*}[h!]
  \begin{center}
    \caption{Satellite parameters.} 
    \label{tab:table2}
\begin{tabular}{l|c|c|r}
Parameter	 	& \multicolumn{2}{c|}{Value(s)}				& Units \\
\hline
\hline
Mass 			& 1	& 10									& $\Mio$ \\
\hline
Radius 			& 1	& 2.15								& $\Rio$ \\
\hline
$k_{2, \mathrm sat}$	& \multicolumn{2}{c|}{0.299}					& \\
\hline
$\Delta t_{\rm sat}$ 	& \multicolumn{2}{c|}{0.0, 0.01, 0.1, 1.0} 	& $\Delta t_{\oplus}$ \\
\hline
$\sigma_{\rm sat}$(for $\Delta t_{\rm sat}  = \Delta t_{\oplus}$) 	& $5.41\times 10^{17}$	& $1.17\times 10^{16}$	& $\Msun^{-1}$.AU$^{-2}$.day$^{-1}$\\
\hline
    \end{tabular}
    \tablefoot{Here, we consider two masses for the satellites, $1~\Mio$ and $10~\Mio$. The values of $\sigma_{\rm sat}$ are  calculated for $\Delta t_{\rm sat} = 1.0 \times \Delta t_{\oplus}$.}
  \end{center}
\end{table*}
 
\begin{figure}[h]
\centerline{\includegraphics[width=\columnwidth]{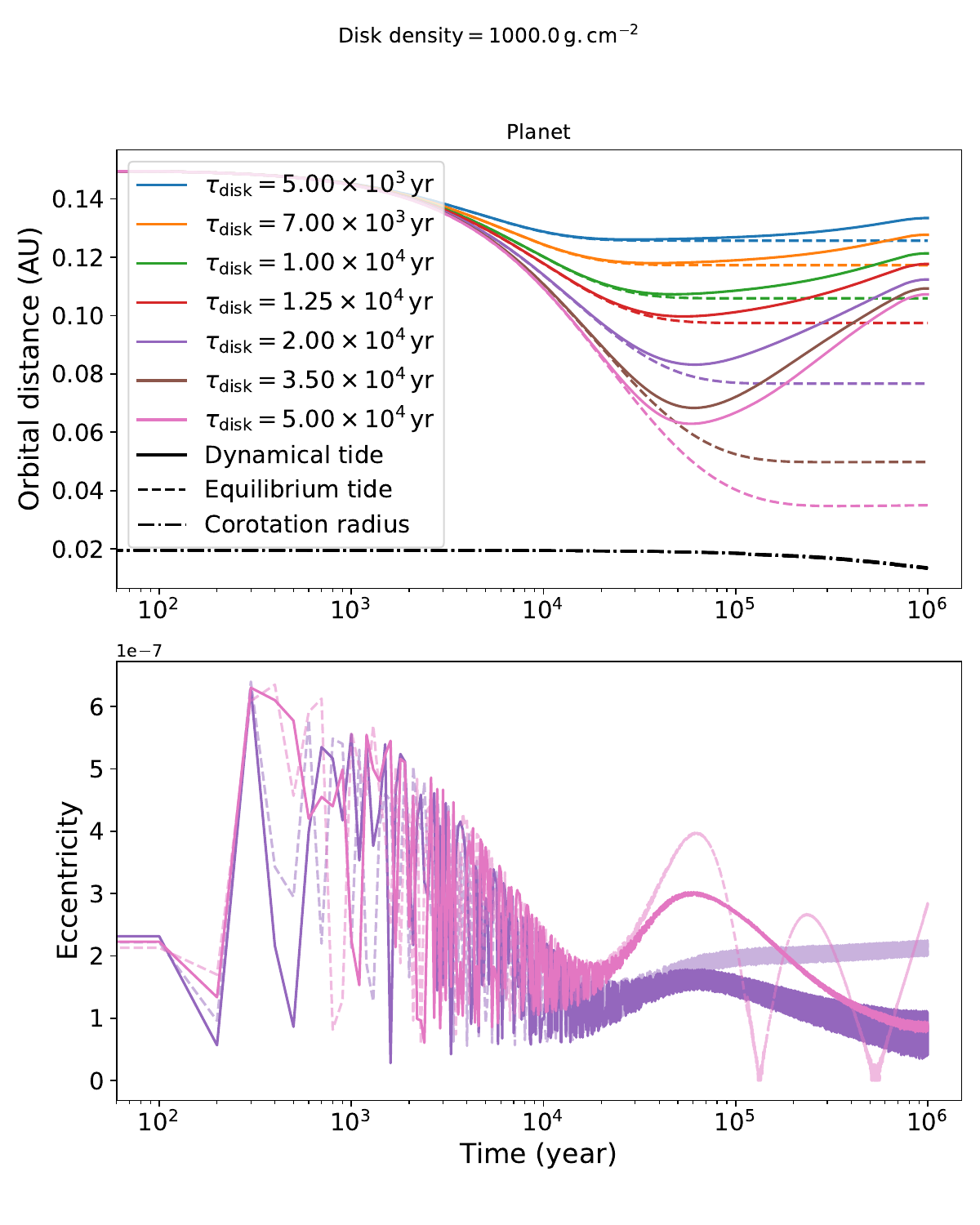}}
\caption{Evolution of the semi-major axis and eccentricity of a Jupiter-mass planet due to type-II migration for different disk lifetimes $\tau_{\rm disk}$ and different assumptions on the stellar tide. The dashed lines corresponds to the case where only the equilibrium tide in the star is accounted for, the solid lines corresponds to cases where both the equilibrium tide and dynamical tide are taken into account (initial fast stellar rotation period of 1~day). {Note that the rotation of the star is evolving, here mainly due to the contraction of the radius, and its evolution is encoded in the corotation radius curve (i.e. where the orbital frequency is equal to the stellar spin). For clarity, we only plot 2 cases in the bottom panel ($\tau_{\rm disk} = 2\times10^4$~yr and $5\times10^4$~yr). Equilibrium tide is still represented with dashed lines but also with transparency to ease identification.}}
\label{Fig4}
\end{figure}

To analyze the impact of type-II migration on a Jupiter-mass planet, we run a \textsc{Posidonius} simulation where the planet begins its evolution at 0.15 AU around a 1~Myr old star {(Figure~\ref{Fig4})}.
The simulation accounts for four key processes: 1) disk interactions, 2) tidal interactions, 3) stellar evolution, and 4) planetary evolution. 
The top panel of {Figure~\ref{Fig4}} (dashed-lines) illustrates how the migration process depends on the disk lifetime $\tau_{\rm disk}$. 
{The results corresponding to the dashed lines were calculated for weak tides (equilibrium tides, see next Section~\ref{tidal_model}), which is very similar to what we obtain with no tides at all.}

A shorter disk lifetime results in migration halting at a greater distance from the star.
For a duration of $\tau_{\rm disk}=5000$~yr, the planet stops at $\approx0.13~$AU, whereas for the longest duration considered of $\tau_{\rm disk} = 50000$~yr, the planet comes as close as $\approx0.03~$AU. 
{The disk lifetime can therefore be used as a proxy for final planet position.}

\subsection{Tidal model}\label{tidal_model} 

To model tidal interactions, \textsc{Posidonius} follows the equilibrium tide prescription of \citet{2015A&A...583A.116B}, implementing the constant time lag model \citep[][]{1979M&P....20..301M,1981A&A....99..126H,1998ApJ...499..853E}. 
The governing equations are detailed in \citet{2020A&A...635A.117B}, where \textsc{Posidonius} was applied to the TRAPPIST-1 system \citep{2017Natur.542..456G}.
Although it features an implementation of the constant time lag model, we adopt different assumptions for the three types of bodies in this study (a satellite, a planet and a star).
{In the following, regardless of how the tidal dissipation is expressed (whether as a quality factor $Q$, a time lag $\Delta t$, or a tidal dissipation factor $\sigma$), the satellite, planet, and star will be denoted by the subscripts $\mathrm{sat}$, $\mathrm{p}$, and $\star$, respectively.}
The specific assumptions and models are outlined below.

\subsubsection{Equilibrium tide in the satellite}

For the satellite, we use a simple constant time lag model, where the dissipation is governed by the quantity $k_{2, \mathrm sat} \Delta t_{\rm sat}$, where $k_{2, \mathrm sat}$ is the satellite's Love number of degree 2, and $\Delta t_{\rm sat}$ is its time lag (assumed to be independent of the excitation frequency in this model). 
As a reference, we use as Earth's time lag value of  $\Delta t_{\oplus} = 638~$s \citep{1997A&A...318..975N}.
\textsc{Posidonius} expresses dissipation through the factor $\sigma$, as introduced in \citet{1998ApJ...499..853E}
\begin{equation}\label{sigma}
\sigma_{\rm sat} = \frac{2}{3} \frac{\G k_{2, \mathrm{sat}}\Delta t_{\rm sat}}{\Rsat^5}, 
\end{equation}
where $\Rsat$ is the satellite's radius.
The corresponding values used in this study are listed in Table~\ref{tab:table2} for the two satellites masses considered: $1~\Mio$ and  $10~\Mio$.
For the $10~\Mio$ satellite, the radius is computed assuming a density equal to that of Io's.

\subsubsection{Equilibrium tide and dynamical tide in the star}\label{subsec:eqdyntidestar}

To model tidal interactions in the star, we adopt the formalism of \citet{2016CeMDA.126..275B}, which is based on the work of \citet{2013MNRAS.429..613O} and \citet{2015A&A...580L...3M}. 
This approach allows us to retain the constant time lag model while incorporating the contribution of the dynamical tide in the specific case of circular and coplanar orbits.
In this context, the dynamical tide refers to the tidal inertial waves that can be excited in the star's convective region.
These waves are driven by the Coriolis acceleration, while the dissipation results from turbulent friction exerted by convective eddies on the waves.
These waves are excited for a very specific range of excitation frequencies $\omega$, which depends on the star's rotation frequency as is defined by $\omega \in [-2\Os, 2\Os]$.
Within this range, the dynamical tide introduces additional dissipation beyond the classical equilibrium tide.
Outside of $[-2\Os, 2\Os]$, only the equilibrium tide contributes to the system’s tidal evolution.
For the equilibrium tide, we assume a dissipation of $\sigma_\star = 4.992\times10^{-66}$~g$^{-1}$cm$^{-2}$s$^{-1}$ following \citet{2010ApJ...723..285H} and \citet{2015A&A...583A.116B}.
  
\citet{2013MNRAS.429..613O} showed that the frequency-averaged dissipation of these tidal waves depends on the structural parameters of the star (radius, mass and radius aspect ratio) and its rotation.
As the star evolves (shrinking or expanding, developing a radiative core, and experiencing changes in rotation), its tidal dissipation also evolves \citep{2015A&A...580L...3M}.
In particular, \citet{2016CeMDA.126..275B} found that this additional dissipation, being significantly stronger than the usual equilibrium tide dissipation, leads to a drastically different early evolution of massive planets when the dynamical tide is accounted for.
 
This formalism was incorporated in \textsc{Posidonius} using the expression for the time lag $\Delta \tau_\star$, as given in \citet{2016CeMDA.126..275B}.
{The following expression is valid for a circular coplanar orbit, for which the excitation frequency is $\omega = 2(n-\Os)$}
\begin{equation}\label{timelag}
\Delta \tau_\star = \frac{3\hat{\epsilon}_\star^{2}}{4k_{2,\star}Q'_{s,\star} |n-\Os|}
\end{equation} 
where $k_{2,\star}$ the Love number of the star, and $Q'_{s,\star}$ represents the frequency-averaged of the structural dissipation due to the tidal inertial waves excited in the star's convective envelope. 
Here, we compute the frequency-averaged structural dissipation factor $Q'_{s,\star}$ from stellar evolution models \citep{2017A&A...604A.112G}, using the following formula
\begin{eqnarray}\label{stellar_freq_av_diss}
\lefteqn{\left<{\mathcal D_\star}\right>_{\omega}=\int^{+\infty}_{-\infty} \! {\rm Im} \left[k_2^2(\omega)\right] \,\frac{\mathrm{d}\omega}{\omega} = \hat{\epsilon}_\star^{2} \frac{3}{2Q'_{s,\star}} }\\
&& = \hat{\epsilon}_\star^2 \left(\frac{\Rs}{\Rsun}\right)^3\left(\frac{\Msun}{\Ms}\right) \frac{100 \pi}{63} \left(\frac{\alpha_\star^5}{1-\alpha_\star^5}\right)\left(1-\gamma_\star\right)^2 \left(1-\alpha_\star\right)^4\\
&&\times\left(1+2\alpha_\star+3\alpha_\star^2+\frac{3}{2}\alpha_\star^3\right)^2\left[1+\left(\frac{1-\gamma_\star}{\gamma_\star}\right)\alpha_\star^3\right]\nonumber\\
&&\times\left[1+\frac{3}{2}\gamma_\star+\frac{5}{2\gamma_\star}\left(1+\frac{1}{2}\gamma_\star-\frac{3}{2}\gamma_\star^2\right)\alpha_\star^3-\frac{9}{4}\left(1-\gamma_\star\right)\alpha_\star^5\right]^{-2}\nonumber
\end{eqnarray}
with
\begin{equation}\label{param_freq_av_diss}
\alpha_\star=\frac{R_{\rm c}}{R_{\star}}\hbox{,}\quad\beta_\star=\frac{M_{\rm c}}{M_{\star}}\quad\hbox{and}\quad\gamma_\star=\frac{\alpha_\star^3\left(1-\beta_\star\right)}{\beta_\star\left(1-\alpha_\star^3\right)}<1.
\end{equation}
The dependence of the dissipation on the stellar rotation is encapsulated in the $\hat{\epsilon}$ parameter, defined as
\begin{equation}\label{epsilon_star}
\hat{\epsilon}_\star = \left( \Os/\sqrt{\G\Msun/\Rsun^3}\right) = \Os/\Omega_{\odot, {\rm c}}, 
\end{equation}
where $\Omega_{\odot, {\rm c}}$ is the critical angular velocity of the Sun. 
As with the satellite dissipation factor in Eq.~\ref{sigma}, the stellar dissipation factor $\sigma_\star$ is derived from $\Delta \tau_\star$, the Love number of the star $k_{2,\star}$, and the radius of the star $\Rs$.

To validate our implementation of the dynamical tide, we reproduced the results of \citet{2017A&A...604A.113B} for a Jupiter-mass planet orbiting a Sun-like star with solar metallicity, considering three different initial stellar rotation periods: 1, 3 and 8~days {(see Appendix~\ref{App_validation}).}
Figure~\ref{Fig4} illustrates the evolution of the semi-major axis and eccentricity of a Jupiter-mass planet, accounting for both the migration in the disk and tidal evolution.
Different line styles represent different assumptions: 
dashed lines correspond to a scenario where only the equilibrium tide is taken into account, while solid lines depict a case where both equilibrium and dynamical tides are included. 
When the additional dissipation of the dynamical tide is considered, the planets do not migrate as close to the star because the stellar tide efficiently pushes them outward, counteracting the  disk-driven migration. 
As a result, a migrating Jupiter-mass planet orbiting an initially fast rotating Sun-like star is unlikely to be found very close to its host star within the first few Myrs of its lifetime.
The bottom panel of Fig.~\ref{Fig4} also tracks the planet's eccentricity, which is initially set to zero. 
In a numerical integration such as those performed by \textsc{Posidonius}, a small residual eccentricity can persist due to numerical artifacts \citep[see][]{2015A&A...583A.116B}, as seen in Fig.~\ref{Fig4}. 
During the first 10'000~yr of evolution, this eccentricity is damped by the disk. As the disk dissipates, a slight excitation of the eccentricity occurs.  
When the dynamical tide is included (solid lines), the stellar tide subsequently damps the eccentricity, an effect that is more pronounced for the {planets closer to the star}.
Conversely, if only the equilibrium tide is considered (dashed line), the eccentricity is not as strongly damped; however, its value remains well under $10^{-6}$.

In our application to a star-planet-satellite system, the orbits are not perfectly circular. 
{The eccentricity of the planet remains small (see Figs. ~\ref{Fig4} and \ref{Fig5}), which means the evolution of the planet due to the stellar tide should be as correct as our formalism allows.
However, the eccentricity of the satellite can be high.
In that case, other excitation frequencies are excited in addition to the main semi-diurnal frequency $\omega=2(n-\Os)$, which would enhance dissipation within the planet.
Taking into account these additional modes might lead to shorter migration and eccentricity damping timescales for the satellite.}
As a result, we are using a prescription that technically falls outside its formal applicability limits.
However, implementing a more accurate formalism for the star would introduce significant complexity.
A proper treatment would require abandoning the convenient, simple, and computationally efficient constant time lag model in favor of accounting for the frequency dependence of tidal dissipation in stars and planets by implementing the Kaula formalism \citep{1964RvGSP...2..661K}. 
These improvements have already been implemented in \textsc{Posidonius} for the planets by \citet{2024A&A...691L...3R}, and are currently being implemented for stars (Kwok et al. in prep). 
Given the planet eccentricity remains small and the dynamical tide has a non-negligible effect on close-in planets \citep{2016CeMDA.126..275B}, we opted to use this formalism despite its limitations.
While not perfect, this approach constitutes an improvement over the original constant time lag model, as it allows the time lag to evolve with the stellar and planetary parameters.

\subsubsection{Equilibrium tide and dynamical tide in the planet}\label{subsubsec:eq_dyn_tid_in_planet}

For the planet, we adopt a similar approach as the for the star.
The gas giant, with mass $\Mp$, radius $\Rp$, and density $\rho_p$, is modeled as a two-layer structure consisting of a homogeneous and a fluid envelope.
The core, with radius $R_c$, mass $M_c$, and density $\rho_c$, is surrounded by a fluid envelope where the dissipation of inertial waves occurs.

For the equilibrium tide, we assume a dissipation value of $\sigma_p = 2.006\times10^{-60}$~g$^{-1}$cm$^{-2}$s$^{-1}$, corresponding to the dissipation of a purely convective body \citep{2012ApJ...757....6H, 2015A&A...583A.116B}\footnote{For a comparison between the various values of dissipation we consider here and the typical tidal quality factor $Q_p$, please refer to Appendix~\ref{App1}.}.

For the dynamical tide, we compute the frequency-averaged structural dissipation factor $Q'_{s,p}$ using data from \citet{2013NatGe...6..347L} and applying the formalism of \citet{2013MNRAS.429..613O} \citep[see also][]{2014A&A...566L...9G}.
The frequency-averaged dissipation due to tidal inertial waves in the fluid envelope of a piecewise-homogeneous fluid body with a solid core of a different density is given by \citet{2013MNRAS.429..613O,2014A&A...566L...9G}
\begin{align}\label{plan_freq_av_dissip}
\left<{\mathcal D_{p}}\right>_{\omega} &=\int^{+\infty}_{-\infty} \! {\rm Im} \left[k_2^2(\omega)\right] \,\frac{\mathrm{d}\omega}{\omega} = \hat{\epsilon}_p^{2} \frac{3}{2Q'_{s,p}} \\
 &= \hat{\epsilon}_p^2 \left(\frac{\Rp}{\Rsun}\right)^3\left(\frac{\Msun}{\Mp}\right) \frac{100 \pi}{63} \left(\frac{\alpha_p^5}{1-\alpha_p^5}\right) \\
 &\quad\times\left[1+\frac{1-\rho_p/\rho_c}{\rho_p/\rho_c}\alpha_p^3\right]\left[1+\frac{5}{2}\frac{1-\rho_p/\rho_c}{\rho_p/\rho_c}\alpha_p^3\right]^{-2},
\end{align}
where $\alpha_p$ is the planet's radius aspect ratio, defined as $\alpha_p= R_c/\Rp$, and $\hat{\epsilon}_p$ is set with respect to the Sun as for the stellar tide\footnote{This formulation allows the same computational implementation to handle both planetary and stellar dynamical tides.}
\begin{equation}\label{epsilon_plan}
\hat{\epsilon}_p = \left( \Op/\sqrt{\G\Msun/\Rsun^3}\right) = \Op/\Omega_{\odot, {\rm c}}.
\end{equation}
Note that Equation~\ref{plan_freq_av_dissip} differs from Equation~\ref{stellar_freq_av_diss}, where the core was assumed to be fluid rather than rocky.

Following \citet{2014A&A...566L...9G}, we assumed a core mass of $6.35~\Mearth$ and of radius $1.38~\Rearth$ to calculate $Q'_{s,p}$.
Core mass and radius are taken to be constant while the radius of the planet $\Rp$ evolves as the planet ages \citep[following][]{2013NatGe...6..347L}.

\section{Numerical set-up and initial conditions}\label{setup}

In this study, we consider a Jupiter-mass planet orbiting a Sun-like star while hosting a satellite.
The initial age of the Jupiter-mass planet and the star is set to 1~Myr.
The initial rotation periods of the planet and the satellite are 13 and 10 hours, respectively, with both bodies having an initial obliquity of 0.2 rad.

The parameters varied in our simulations include: 
\begin{itemize}
\item[-] the initial semi-major axis of the satellite (0.3, 0.4, 0.6, 0.8, 1.0 $\aio$),
\item[-] the mass of the satellite (1~$\Mio$, 10~$\Mio$), 
\item[-] the dissipation in the satellite ($[0,~10^{-4},~10^{-2},~10^{-1},~1.0]\times \Delta t_{\oplus}$), 
\item[-] the disk lifetime (5'000, 7'000, 10'000, 12'500, 20'000, 35'000, 50'000~yr), 
\item[-] the initial stellar rotation period (1~day, 3~days, 8~days).
\end{itemize}

Since our primary focus is on the final stages of the planet migration, we initialize the simulations with a planet at 0.15 AU from the star.
{At that initial star-planet distance, the outer satellite (1.0 $\aio$) is initially well within the 0.5~R$_{\mathrm{Hill}}$ stability criterium \citep{2006MNRAS.373.1227D} and we note a limited eccentricity excitation ($\lesssim 10^{-2}$).}
This also justifies the relatively short disk lifetime compared to the typical migration timescales required to explain the inward migration of a Jupiter-like planet from 5 AU to a few $10^{-2}$~AU.
{The integration timescale is of 5-6 Myr, for the simulations with surviving satellites.
For the others, the integration is stopped when the satellites collides with the planet.
Note that due to long computation time for the closest satellite-planet distances, limited resources and overall footprint of these calculations, some simulations were not continued to reach 5 Myr.}

\section{Between a rock and hard place}

The survival of the satellite is constrained by two key processes: tidal inward migration and eccentricity excitation, both of which can ultimately lead to a collision. 
The relative significance of these processes depends on two main factors: 1) the disk lifetime and density, which serve as proxies for the final star-planet distance resulting from the migration, and 2) the dissipation properties of the various bodies involved. 
We discuss these influences in the following sections. 
However, before getting into a broader analysis, we first illustrate the possible outcomes through a specific case in which only the equilibrium tide in the star and planet is considered. 

\subsection{Equilibrium tide in star and planet}\label{Results_Eq_tide}

Since most tidal studies consider only the equilibrium tide, we begin our analysis with this familiar set-up.
In this section, we first discuss the influence of the satellite's initial semi-major axis (Section~\ref{subsec:initial_distance}), followed by the effects of disk lifetime (Section~\ref{subsec:disk_lifetime}), satellite dissipation, and satellite mass.

\subsubsection{Initial distance of the satellite}\label{subsec:initial_distance}

\begin{figure*}[h]
\centerline{\includegraphics[width=\linewidth]{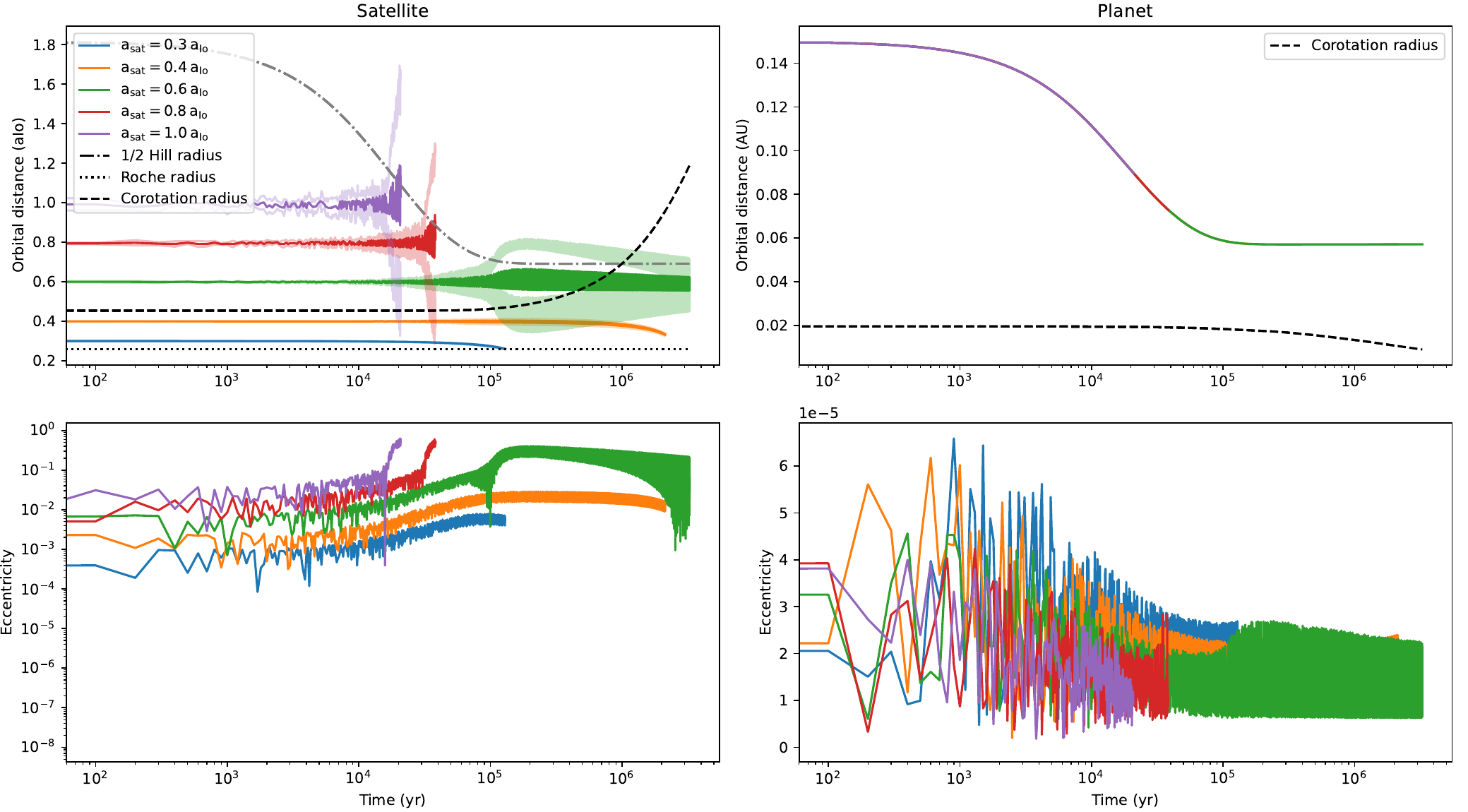}}
\caption{Evolution of a $1~\Mio$ satellite orbiting a migrating Jupiter-mass planet for different initial distances of the satellite. Here, the disk lifetime is $3\times10^4$~yr, the initial rotation period of the star is 1~day and the dissipation in the satellite is taken to be 0. The top left panel shows the evolution of the semi-major axis of the satellite (full colored lines), the shaded colored lines represent the periastron and apoastron. The gray dotted line represents the radius of the planet, the black dash-dotted line represents the Roche radius, the black dashed line the corotation radius (where the satellite's mean motion is equal to the spin of the planet) and the grey dashed line represents the half of the Hill radius of the planet. The bottom left panel shows the evolution of the satellite's eccentricity. The top right panel shows the evolution of the semi-major axis of the planet (full colored lines). The black dashed line represents the corotation radius (where the planet's mean motion is equal to the spin of the star). The bottom left panel shows the evolution of the planet's eccentricity.}
\label{Fig5}
\end{figure*}

The evolution of a $1~\Mio$ satellite orbiting a migrating Jupiter-mass planet varies depending on its initial orbital distances, as illustrated in Figure~\ref{Fig5}. 
In this case, we assume a disk lifetime of $3\times10^4$~yr.
Under these conditions, the planet reaches a final distance of 0.057~AU after slightly more than 100'000~years of migration (top right panel of Fig.~\ref{Fig5}).
{We can therefore consider here the disk to be fully dissipated at around 100'000~years after the beginning of our simulation.}
Throughout the migration, the planet's eccentricity remains low, primarily due to the damping effect of the disk (bottom right panel of Fig.~\ref{Fig5}).
As the planet approaches the star in the final stage of the migration, its rotation evolves in response to the shifting corotation radius, as shown in the top left panel (black dashed lines). 
Due to the tide raised by the star in the planet, the planet's rotation gradually slows down towards the pseudo-synchronous rotation, which, for such low eccentricities, closely approximates synchronous rotation (see bottom right panel). 
In this specific case, the final pseudo-synchronous period is approximately 5~days {and the corresponding plot can be seen in Appendix~\ref{FigApp_rotation_planet}}.

In the scenario depicted in Fig.~\ref{Fig5}, we assume no dissipation in the satellite.
The satellite's fate varies depending on its initial orbital distance\footnote{{Note that not all satellites are integrated until they either fall onto the star or reach 6 Myr. This is due to the fact that the closer the satellite the smaller the integration timestep. So for the cases where the fate is determined like the orange satellite, or where the integration to 6~Myr will not bring additional insights, we did not continue the integration.}}. 
First, consider a satellite that begins its orbit beyond 0.8~$\aio$ (the two outermost satellites). 
As the planet migrates inward, the increasing gravitational influence of the star excites the satellite’s eccentricity. 
This effect is illustrated by the grey {dot-dashed} line, which represents half of the Hill radius, R$_{\mathrm{Hill}}$, of the planet.
This follows the stability limit proposed by \citet{2006MNRAS.373.1227D}, who suggested that satellites remain stable within approximately $\sim 0.5$~R$_{\mathrm{Hill}}$. 
The Hill radius marks the distance at which the satellite experiences equal gravitational attraction from both the star and the planet. 
As the planet moves closer to the star, the Hill radius shrinks, intensifying the eccentricity excitation. 
For the two outermost satellites, this excitation is strong enough to bring their periastron distance within the Roche radius of the planet. 
As a result, the satellite undergoes tidal disruption, effectively leading to a collision with the planet.
We refer to this outcome as {dynamical disruption}, distinguishing it from the alternative fate described next, which occurs under different migration conditions. 

Next, we consider the fate of the innermost satellite. 
The satellite initially located at 0.3~$\aio$ undergoes inward migration due to the tide raised by the satellite in the planet, as we have assumed no dissipation in the satellite itself. 
This migration has two contributing factors: 1) since the satellite orbits below the corotation radius (represented as a black dashed line), the planetary tide acts to decrease the semi-major axis; 2) the satellite's non-zero eccentricity induces eccentricity-driven tidal migration. 
{Following an initial stage of eccentricity excitation due to the approaching star, the eccentricity is then damped and the semi-major axis also decreases (blue, orange and green curves).}
It is important to note that, in this case, eccentricity damping is driven solely by tides raised inside the planet, which is significantly less efficient than the dissipation that would occur if tides were raised in the satellite (a process explored later in this study).
{Due to the satellite's relatively low mass, the migration induced by the tide raised in the satellite (contributing factor 2 above) dominates its evolution.}
Moreover, this effect is further amplified by an additional mechanism: as the star approaches, it triggers eccentricity excitation, preventing the eccentricity from fully damping to zero.
This eccentricity leads to a slightly faster migration compared to a scenario with zero-eccentricity. 
For the innermost satellite, this tidally-induced inward migration progresses rapidly enough that it reaches the Roche radius of the planet in about 100'000~yr, resulting in tidal disruption. 
We refer to this outcome as {tidal disruption}, distinguishing it from the previously described {dynamical disruption}, as the underlying mechanism differs.  

The satellite initially located at 0.4~$\aio$ follows a very similar evolutionary path, though it does not collide within the simulation time.
Due to its greater initial distance from the planet, its inward migration progresses more slowly, while its eccentricity excitation is slightly more pronounced. 
Although the satellite remains intact within the simulation period, it is likely to reach the Roche radius in a few million years. The exact timescale for this eventual disruption depends on the dissipation assumed for the planet.

The satellite initially located at 0.6~$\aio$ follows an intermediate evolutionary path. 
It undergoes significant eccentricity excitation due to the approaching star, and this enhanced eccentricity leads to some tidal evolution. 
In particular, the planetary tide dampens the eccentricity, leading to a {very slow decrease of the semi-major axis (note that during this phase, due to angular momentum conservation, the pericenter distance does not decrease)}. 
Initially, the satellite orbits beyond the corotation radius, where the planetary tide acts to increase its semi-major axis. 
However, because the satellite is not massive enough, this outward effect remains negligible. 
At approximately 1.5~Myr, the satellite crosses the corotation radius, after which both migration mechanisms (one driven by the planet's non-synchronous rotation and the other by the satellite's non-zero eccentricity) begin working in the same direction: inward migration.
{However, this process is quite long. 
The eccentricity previously excited by the approaching star will be damped until it reaches an equilibrium between tidal damping and eccentricity excitation \citep[e.g.][]{2009ApJ...704L..49B,2013A&A...556A..17B}. 
Once this equilibrium is reached, the satellite will experience a satellite-tide induced inward migration similar to the satellite at 0.3 and $0.4~\aio$ (in blue and orange), enhanced by the slow inward migration due to the weak planetary tide.}
From this point forward, the satellite's fate is sealed, and it will inevitably reach the Roche radius.
Determining the exact moment of its disruption would require a longer integrating time, becoming very computationally expensive.

In summary, in this particular example, no satellite is expected to survive the evolution. 
They are either {dynamically disrupted} early on due to the eccentricity excitation caused by the approaching star, or they are (or will be) {tidally disrupted} following a tidally-induced inward migration. 

\begin{figure*}[h]
\centerline{\includegraphics[width=\linewidth]{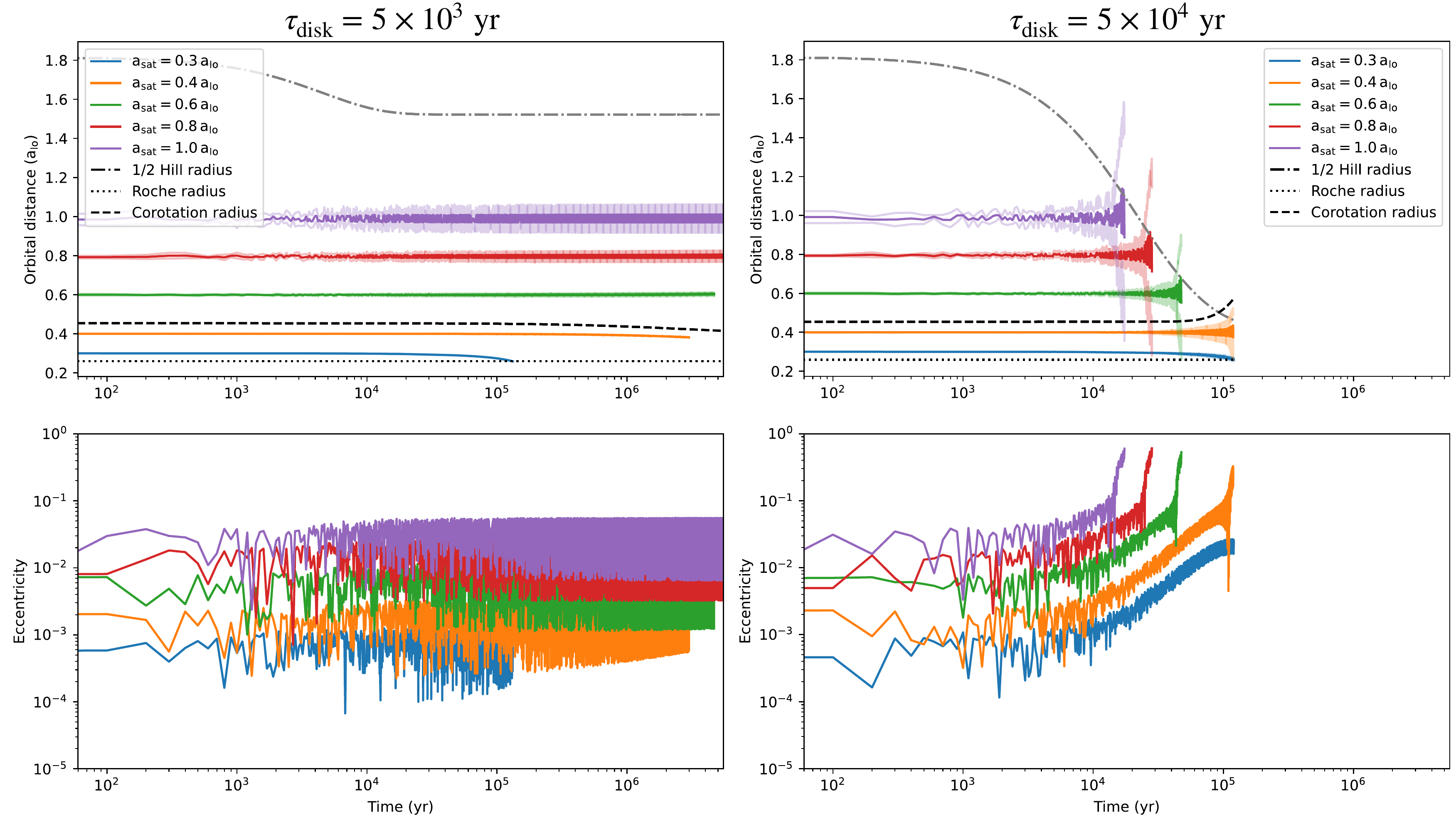}}
\caption{Same as the right column of Figure~\ref{Fig5} but for a disk lifetime of $5\times10^3$~yr (left column) and $5\times10^4$~yr (right column). The top row shows the evolution of the orbital distance of the satellites and the bottom row shows the evolution of their eccentricities. For the left column, the final semi-major axis of the planet is 0.125~AU and for the right column, it is 0.035~AU. This difference can be visualized thanks to the grey dashed lines corresponding to half the Hill radius of the planet, which comes much closer to the planet when the $\tau_{\rm disk} = 5\times10^4$~yr.}
\label{Fig6}
\end{figure*}

\subsubsection{Disk lifetime - or planet final distance}\label{subsec:disk_lifetime}

The disk lifetime plays a crucial role in determining the survival of the satellites. 
A longer disk lifetime allows the planet to migrate closer to the star, increasing eccentricity excitation and thereby raising the likelihood of the satellite colliding with the planet. 

To illustrate the effect of disk lifetime, we compare the evolution of the satellite for the two extreme values of the disk lifetime considered: $5\times10^3$~years in (left column of Figure~\ref{Fig6}) and $5\times10^4$~years (right column of Figure~\ref{Fig6}).
As previously discussed, the disk lifetime $\tau_{\rm disk}$ serves as a proxy for the planet's final position relative to the star.  
In the previous section, with $\tau_{\rm disk}$ = $3\times10^4$~yr, the planet completed its migration at approximately 0.057~AU. 
For a shorter disk lifetime of $\tau_{\rm disk}$ = $5\times10^3$~yr, the planet settles farther from the star at 0.125~AU.
Conversely, with a longer disk lifetime of $\tau_{\rm disk}$ = $5\times10^4$~yr, the planet migrates inwards to 0.035~AU.

In the case of the shortest disk lifetime, where the planet settles at 0.125~AU, most satellites appear to survive the evolution. 
Only the two innermost satellites have either undergone tidal disruption or seem to be evolving towards it.
The fate of the satellite initially located at 0.4~$\aio$ remains uncertain. 
The slight decrease in the corotation radius suggests that this satellite could eventually cross it, following a process similar to that described in \citet[][]{2011A&A...535A..94B}. 
If this occurs, the eccentricity-driven inward migration could be counteracted by the outward migration caused by the non-synchronicity between planet's rotation rate and the satellite's orbital frequency.
The gradual decrease in corotation is a result of planetary spin-up induced by the planet's contraction. 
Because the planet remains relatively distant from the star, this spin-up occurs on shorter timescale than the tidally-induced spin-down, which will eventually slow the planet's rotation to approximately 15~days {(corresponding to a corotation radius increase to about $4~\aio$)}.
Although longer simulations would be required to determine the satellite's precise evolution, it is {thus} expected to eventually cross the increasing corotation radius, undergo inward migration, and ultimately reach the Roche limit.   
{However, for simulations of satellites that are very close to the planet, the timestep is very small, which means that the simulations are very long. 
The full computation of the evolution of this satellite would require an unreasonable amount of resources.}

In the case of the longest disk lifetime, where the planet settles at 0.035~AU, all the satellites are ultimately lost. 
The four outermost satellites undergo dynamical disruption due to the eccentricity excitation induced by the approaching star, while the innermost satellite is tidally disrupted.
The planet barely has time to reach its final semi-major axis before all the satellites are removed. 
An interesting consequence of the planet's proximity to the star is the rapid evolution of its rotation towards the pseudo-synchronous rotation. 
This effect is evident in the significant increase in the corotation radius over the course of the first few $10^5$~yr of the simulation.

In this scenario, the tidal evolution of the satellites is solely dictated by the tide they raise in the planet, as we assumed no dissipation in the satellite. 
This represents the most favorable case for the configuration considered here, where only equilibrium tides in the planet and star are taken into account. 
Including tidal dissipation in the satellite would further increase the probability of tidal disruption. 
A synthetic visualization of these results are presented in the top left panel of Figure~\ref{Fig8} (1.0~$\Mio$ and $\Delta t_{\rm sat} = 0.0~\Delta t_\oplus$), where the different populations are depicted. 
Satellites that undergo  tidal disruption are shown in red (those initially located at $a_{\rm sat} = 0.3~\aio$), while those experiencing tidal decay, likely leading to eventual tidal disruption, are marked in orange. 
Dynamical disrupted satellites, which are the farthest from the planet in cases with the longest disk lifetime, appear in blue. 
Surviving satellites are displayed in green.
For these surviving satellites, their evolution follows one of two paths: either they undergo a slight outward migration driven by the planetary tide (as long as they stay beyond the corotation radius), or they exhibit no significant migration.  
The threshold distinguishing non-negligible migration from negligible migration is taken to be $10^{-4}$~AU/Myr, corresponding to approximately 0.035~$\aio$/Myr).  

\begin{figure}[!htbp]
\centerline{\includegraphics[width=\linewidth]{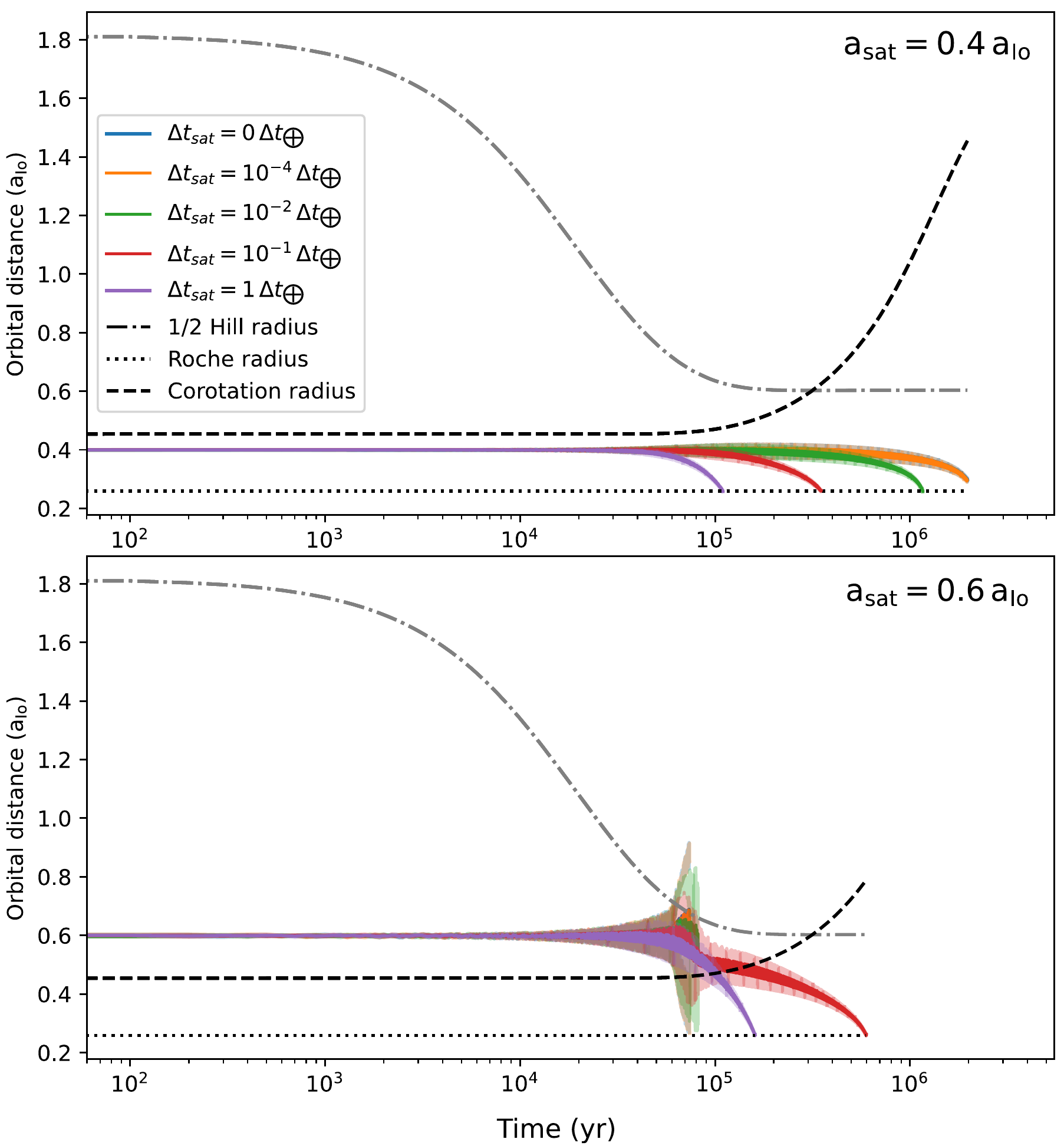}}
\caption{Evolution of the semi-major axis and eccentricity {(through apoastron and periastron distances)} of a 1.0~$\Mio$ satellite for different satellite dissipations. Here, the disk lifetime is 35'000 yr, and the disk density 1000~g.cm$^{-2}$. The evolution is shown for two different initial satellite semi-major axis: 0.4~$\aio$ {(top panel)} and $0.6~\aio$ {(bottom panel)}.}
\label{Fig7}
\end{figure}

The disk lifetime, or equivalently the planet's proximity to its star, is a crucial factor in determining satellite survival. 
As shown in Fig.~\ref{Fig8}, no satellite remains intact if the planet migrates closer than 0.057~AU.
At shorter distances, the outer satellites undergo dynamical disruption, while the inner satellites are likely to experience tidal disruption.
Conversely, if the planet remains beyond 0.057~AU, the outer satellites can survive.

\subsubsection{Increasing the dissipation in the satellite}

As previously discussed, assuming zero dissipation in the satellite represents an idealized scenario for satellite survival. 
In reality, no bodies exhibit zero dissipation, hence we now investigate the impact of varying this parameter. 
To explore a broad range of possibilities, we vary the dissipation from 0 to 1 times the dissipation of the Earth\footnote{For a comparison between the various dissipations considered here and the typical tidal quality factor $Q_\mathrm{sat}$, please refer to Appendix~\ref{App1}.}
Figure~\ref{Fig7} presents the evolution of a 1~$\Mio$ satellite for a disk lifetime of 35'000~yr, considering different values of the satellite dissipation and two distinct initial orbital distances: $0.4~\aio$ (left column) and $0.6~\aio$ (right column). 

The expected effect of increasing dissipation is clearly observed in the top panel of Fig.~\ref{Fig7}: inward migration occurs on shorter timescales. 
As dissipation in the satellite becomes increasingly more effective at damping the excited eccentricity ({slightly visible with the periastron and apoastron distances: the line is thinner for higher dissipations}), migration accelerates, ultimately leading to early tidal disruption. 

The {bottom} panel of Fig.~\ref{Fig7} presents another example, offering additional insights. 
Notably, for the two highest dissipation values (0.1 and 1.0~$\Delta t_\oplus$), the satellite is no longer dynamically disrupted. 
As the star approaches the planet-satellite system, eccentricity excitation increases, but the effectiveness of tidal damping depends on the level of dissipation. 
Lower dissipation results in weaker tidal damping, allowing eccentricity to rise unchecked.
For the lowest dissipation cases, the periastron distance eventually decreases to the Roche radius, leading to dynamical disruption.
Conversely, at the two highest dissipation levels, the satellite tide acts to sufficiently damp the eccentricity to prevent that fate. 
For 0.1~$\Delta t_\oplus$ (in red), eccentricity damping from 0.3 to 0.1 (bottom right panel) accelerates inward migration, which then slows as the eccentricity drops below 0.1. 
For 1.0~$\Delta t_\oplus$ (in purple), stronger dissipation limits eccentricity excitation to values below 0.1 (bottom right panel), resulting in a smoother inward migration to 0.1~$\Delta t_\oplus$.
Once the eccentricity falls below 0.1, both satellites continue to experience a gradual decrease in semi-major axis and eccentricity until tidal disruption occurs before $10^6~$yr.
In that example, the fate of the satellite remains the same: it does not survive, but the dissipation in the satellite changed the nature of its fate from a dynamical disruption to a tidal disruption.
The highest the dissipation, the more satellites experience tidal disruption.

\begin{figure*}[!htbp]
\centerline{\includegraphics[width=0.9\linewidth]{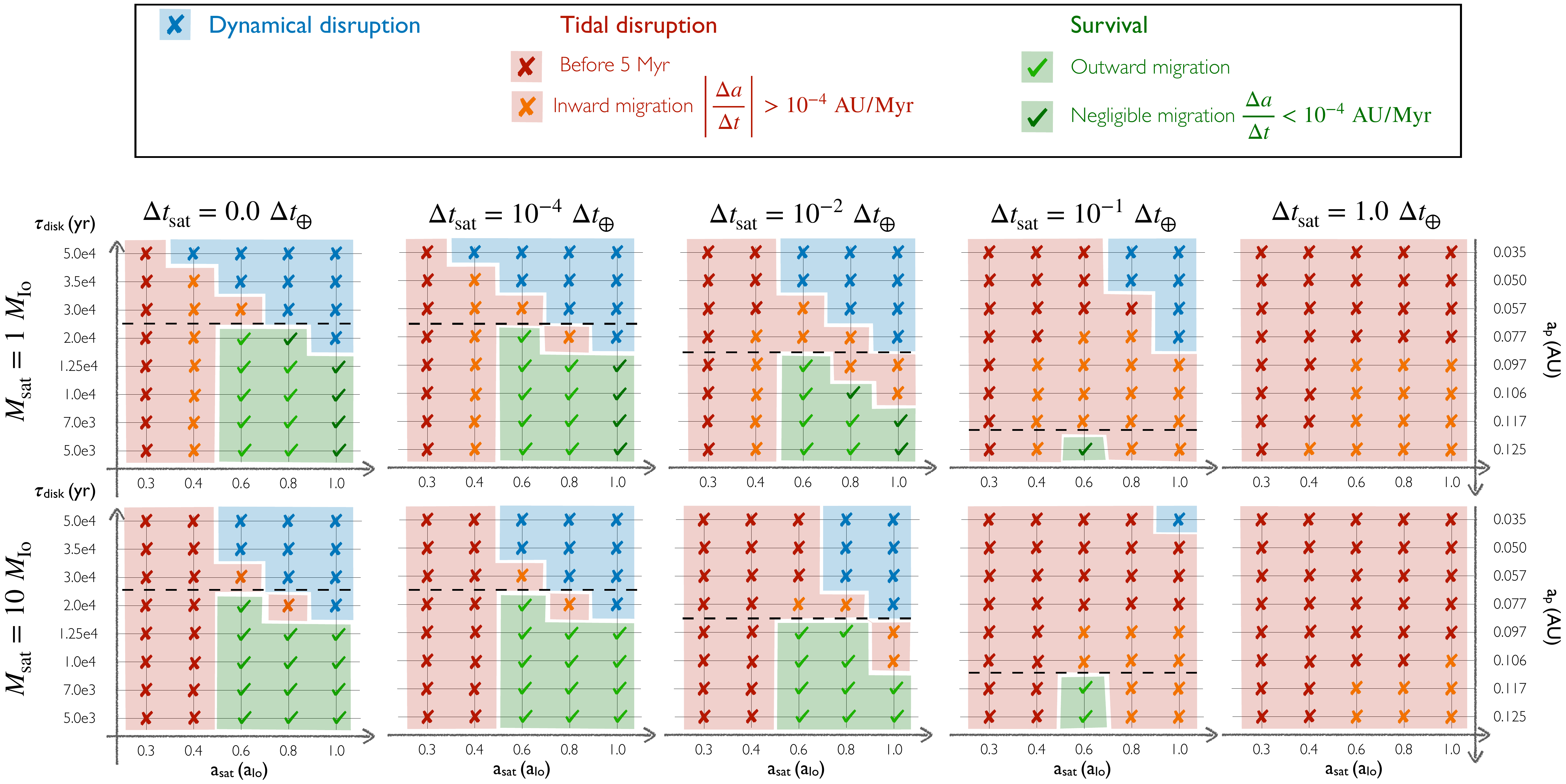}}
\caption{Survival chart of a 1.0~$\Mio$ (first row) and a 10.0~$\Mio$ (second row) satellite for different satellite dissipations {and for equilibrium tides only}: from no dissipation (left) to $\Delta t_{\rm sat} = 1~\Delta t_{\oplus}$ (right). The fate of the satellites is a function of the disk lifetime (left vertical axis) and the corresponding final semi-major axis of the planet (right vertical axis) and its initial position (horizontal axis). The blue crosses and region correspond to the dynamical disruption scenario. The red crosses correspond to satellite that underwent a tidal disruption within the simulation time (5~Myr), the orange crosses correspond to satellites which are migrating inwards at a rate faster than $10^{-4}$~AU/Myr. This population (red and orange crosses) is outlined by the red region. The dark green check signs correspond to satellites whose fate is not clear (migration lower than $10^{-4}$~AU/Myr). The light green check signs correspond to an outward migration at a rate faster than $10^{-4}$~AU/Myr. The green region therefore corresponds to satellites which should survive. The dashed black line shows the final planet semi-major axis below which there is no survival of satellites.}
\label{Fig8}
\end{figure*}

A synthetic view of the satellites' fates across different dissipation values is presented in Figure~\ref{Fig8}, with dissipation increasing from left to right. 
The results indicate that higher dissipation leads to a greater minimum star-planet distance required for satellite survival. 
As dissipation increases from 0 to 0.1~$\Delta t_\oplus$, this survival threshold shifts from 0.077 to 0.125~AU.
For 1~$\Delta t_\oplus$, satellite survival becomes impossible at star-planet distances below 0.125~AU, which is the biggest distance considered in this study.
Thus, to detect a satellite around a planet orbiting closer than 0.1~AU, the satellite's internal dissipation must be relatively low.

Given that the heat dissipation on Io is estimated to be approximately 3~W/m$^2$ \citep[e.g.][]{2000Sci...288.1198S}, we can derive a corresponding $\Delta t_{\rm Io}$ to compare with the dissipation values we used in this study (expressed in $\Delta t_\oplus$). 
Using the formula to compute tidal heating in the framework of the CTL model \citep{2010A&A...516A..64L,2013A&A...556A..17B} and extracting the time lag corresponding to Io's tidal heating yields a value of $\Delta t_{\rm Io} = 2.7\times \Delta t_\oplus$ \citep[consistent with actual values of $\Delta t_{\rm Io}$, see][]{2024GeoRL..5107869A}. 
This value exceeds the maximum $\Delta t_{\rm sat}$ considered in this study.
If Io is representative of such exomoons, this suggests that they are unlikely to exist around planets orbiting closer than 0.125~AU to their star.
At these small separations, eccentricity excitation caused by the star drives the satellite towards tidal disruption and eventual engulfment.

\subsubsection{Increasing the mass of the satellite}\label{Results_Eq_mass_sat}

We now examine the evolution of a more massive satellite with a mass of $10~\Mio$.
A higher satellite mass leads to a faster orbital evolution: it induces stronger tidal interactions with the planet, making its migration more sensitive to its position relative to the corotation radius.
Additionally, due to its larger radius compared to the $1~\Mio$ satellite, it undergoes a more rapid evolution driven by satellite-induced tides.
A synthesized comparison of these effects is presented in Figure~\ref{Fig8}, allowing for a direct evaluation of differences between the $10~\Mio$ and $1~\Mio$ satellites.

For all the disk lifetimes considered in this study, and as expected, satellites initially located inside the corotation radius undergo tidal disruption more rapidly than their $1.0~\Mio$ counterparts. 
This trend is evident in Figure~\ref{Fig8}, all $10~\Mio$ satellites that began at 0.4~$\aio$ experience tidal disruption before 5~Myr (red crosses), whereas their $1~\Mio$ counterparts remain intact (orange crosses).

The fate of satellites initially located outside the corotation radius depends on the disk lifetime under consideration.
First, we examine the shortest disk lifetime (5000~yr), which corresponds to a final star-planet distance of 0.125~AU.  
For satellites with no dissipation, those initially beyond the corotation experience a significantly more pronounced outward migration than their $1.0~\Mio$ counterparts. 
This effect is evident when comparing the green ticks between top and bottom row of Figure~\ref{Fig8}: in the bottom row ($10~\Mio$), the ticks appear brighter, indicating that satellites more frequently migrate outward at a rate exceeding $10^{-4}$~AU/Myr, equivalent to $3.6\times10^{-2}~\aio$/Myr.
As dissipation increases, the extent of the outward migration is progressively reduced.
For a high dissipation (>0.01~$\Delta t_\oplus$), migration is even reversed for the two outer satellites (0.8 and 1.0~$\aio$), whose eccentricities are the most strongly excited by the approaching star, driving them into inward migration instead.

Increasing the disk lifetime (or equivalently, decreasing the final star-planet distance) significantly impacts the survival of $10~\Mio$ satellites, just as it does for $1.0~\Mio$ satellites. 
For the longest disk lifetime considered, no satellite survives. 
However, the mechanisms leading to their destruction differs from those observed for $1.0~\Mio$ satellites. 
In particular, due to their greater mass, $10.0~\Mio$ satellites are less susceptible to dynamical disruption and instead undergo tidal disruption.
The difference is evident in Figure~\ref{Fig8}, where the extent of the blue areas, representing dynamically disrupted satellites, is noticeably smaller for $10~\Mio$ (bottom row) compared to $1~\Mio$ (top row). 

As observed for $1.0~\Mio$ satellites, survival rates are higher at lower dissipation levels. 
However, even a small amount of dissipation is sufficient to limit their long-term survival. 
Satellites initially located outside the corotation radius undergo a more pronounced outward migration, which temporarily spares them from imminent tidal disruption.
However, as they migrate outward, they become increasingly influenced by the gravitational pull of the star. 
This effect enhances their eccentricity, which could ultimately drive their orbit into a slow inward decay, culminating in tidal disruption.

Further details on the influence of the mass of the satellite will be provided in the next section, where we extend our analysis beyond the equilibrium tide and also account for the dynamical tide in both the star and the planet.

\subsection{Dynamical tide in star and planet}\label{Results_Dyn_tide}

\begin{figure*}[!htbp]
\centerline{\includegraphics[width=0.9\linewidth]{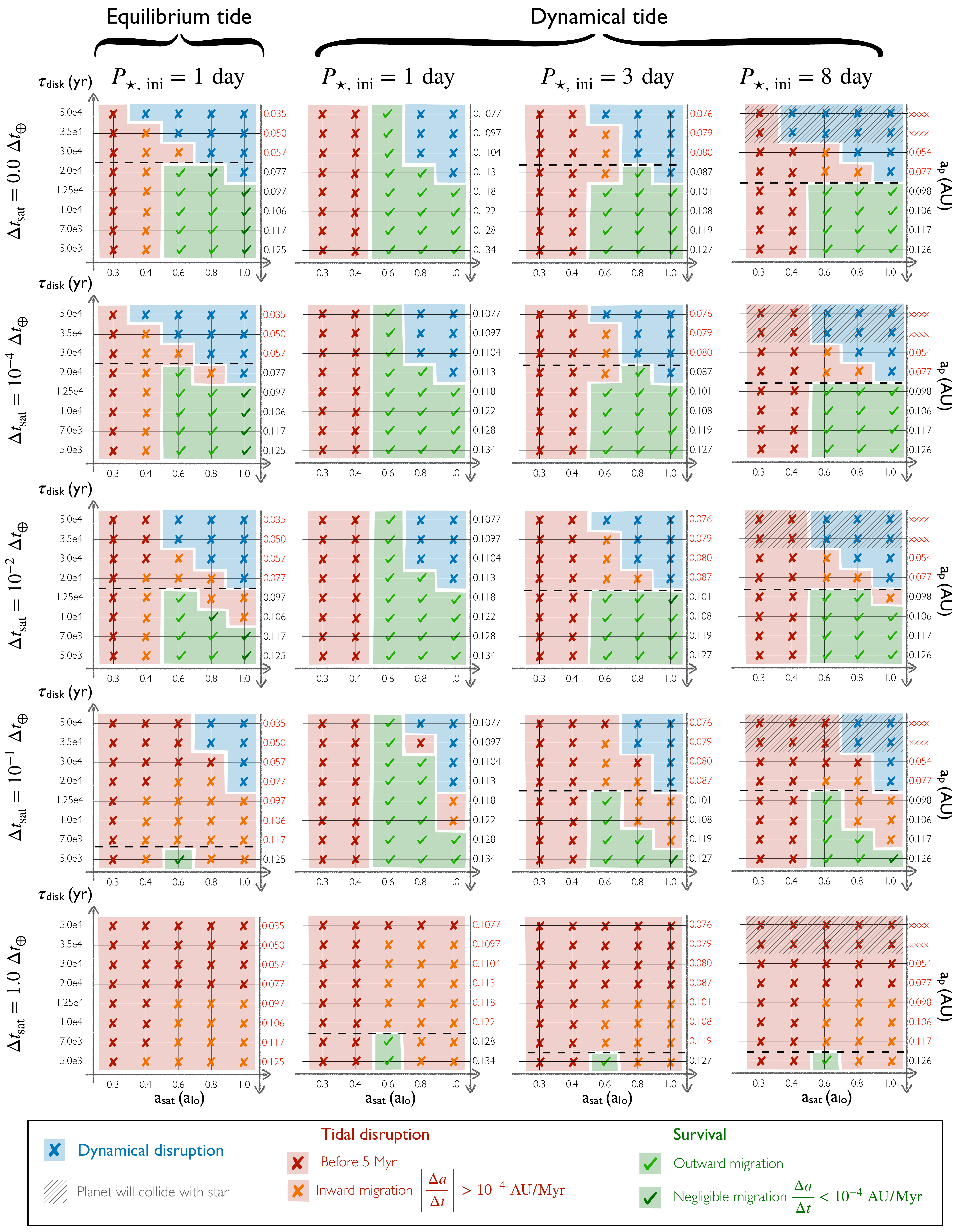}}
\caption{Survival chart of a 1.0~$\Mio$ satellite for different satellite dissipations: from no dissipation (top row) to $\Delta t{\rm sat} = 1.0~\Delta t_{\oplus}$ (bottom row); and for equilibrium tide only (1st column) and dynamical tide (2nd to 4th columns) for different initial stellar rotation periods: 1~day (1st and 2nd columns), 3~day (3rd column) and 8~day (4th column). The fate of the satellites is a function of the disk lifetime (left vertical axis) and the corresponding final semi-major axis of the planet (right vertical axis) and its initial position (horizontal axis). 
The red and blue regions corresponds to where the satellites is disrupted, and the green region where it can survive.
The dashed black line shows the final planet semi-major axis below which there is no survival of the satellites.
The crossed-out area corresponds to cases where the planet itself falls onto the star. 
}
\label{Fig9}
\end{figure*}

In this section, we incorporate the effects of both the dynamical tide in the star and in the planet, in addition to the equilibrium tide\footnote{This approach technically exceeds the formal applicability of our dynamical tide model, which assumes zero eccentricity. However, we argue that it still provides a more realistic approximation than relying solely on the equilibrium tide, as done in Section~\ref{Results_Eq_tide}, by better capturing the enhanced dissipation associated with the dynamical tide}.
When considering only the equilibrium tide in the star, the initial stellar spin has minimal influence. 
However, incorporating the dynamical tide increases dissipation, thereby accelerating the system's tidal evolution, which then becomes highly sensitive on the star's initial rotation rate \citep[see discussions in][]{2016CeMDA.126..275B}.  
Thus, in addition to assuming only an initial rotation of 1~day (representing a fast rotator), we also consider 3~day (an intermediate rotator) and 8~day (a slow rotator). 
A summary of outcomes for 1.0~$\Mio$ satellites is presented in Figure~\ref{Fig9}, illustrating results for different satellite dissipation values (different rows) and the three stellar rotation periods considered (different columns).
To enable direct comparisons with the equilibrium tide case, the first column of Fig.~\ref{Fig9} reproduces the corresponding results from Fig.~\ref{Fig8}.
A similar summary for 10~$\Mio$ satellites is provided in Figure~\ref{Fig12} of {Appendix~\ref{App_10Mio} with the corresponding discussion}.

\begin{figure*}[!htbp]
\centerline{\includegraphics[width=0.9\linewidth]{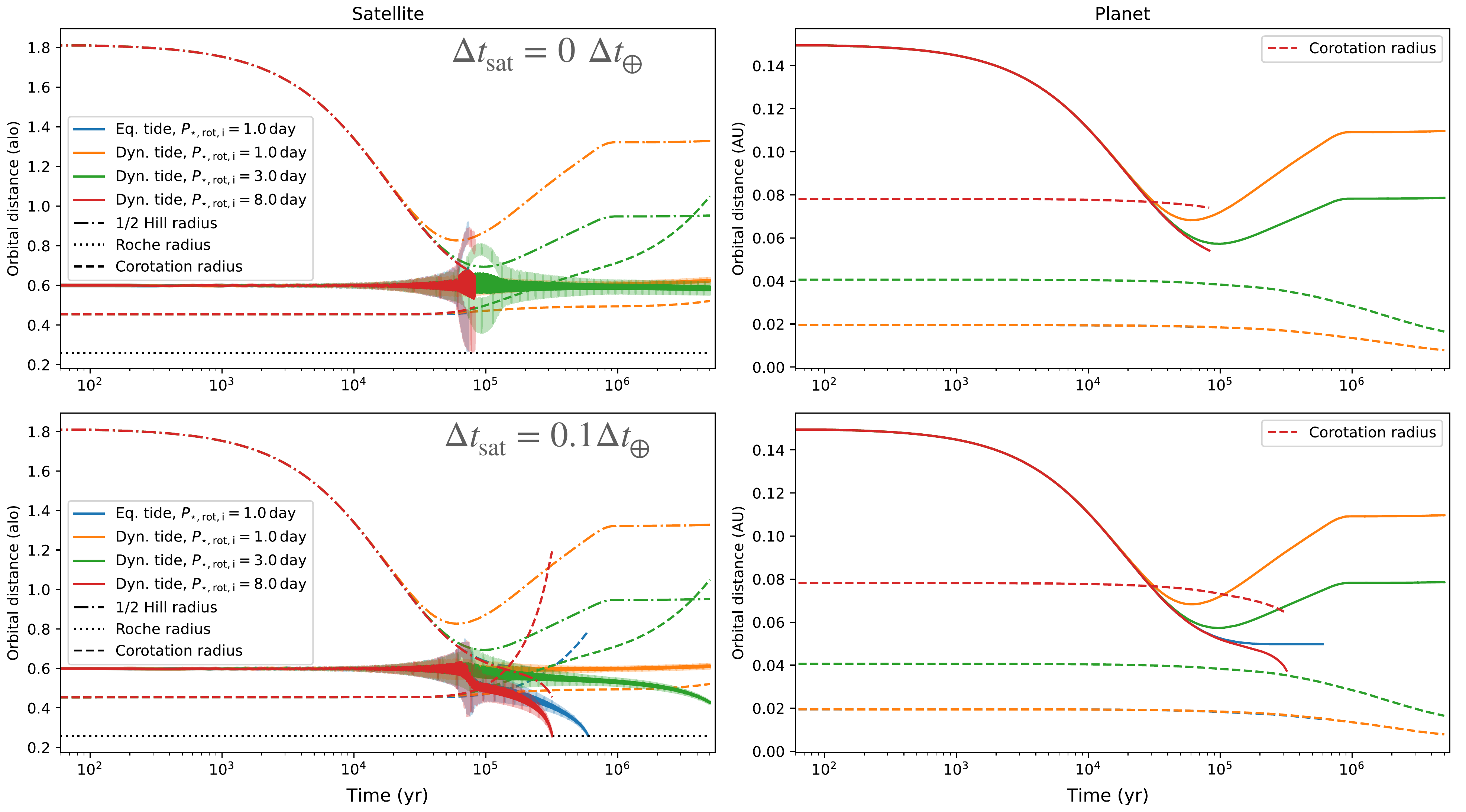}}
\caption{Evolution of the semi-major axis of a 1.0~$\Mio$ satellite (left column) and its planet (right column) for different initial rotation period of the star and 2 different satellite dissipation: $0~\Delta t_{\oplus}$ (top row), and $0.1~\Delta t_{\oplus}$ (bottom row). The disk lifetime is here 35’000 yr, and the disk density 1000 g.cm$^{-2}$.}
\label{Fig10}
\end{figure*}

When accounting for the dynamical tide, the orbital evolution of the planet differs significantly, as shown in Fig~\ref{Fig4} for a fast rotating star. 
Depending on the star's initial rotation period, the planet may be efficiently pushed outwards (for $P_\star = 1$~day) or collide in the star ($P_\star = 8$~day for the two longest disk lifetimes). 
In the latter case, satellite survival is impossible.
Furthermore, incorporating the dynamical tide in the planet enhances dissipation, making planetary-tide driven migration significantly more efficient.
As a result, satellites inside the corotation are engulfed more rapidly than when considering only the equilibrium tide, while those outside corotation migrate outwards at a faster rate than before.

This observation helps explain one of the major differences between the equilibrium tide and the dynamical tide cases in {Fig.}~\ref{Fig9}.
By comparing the maps for the same disk dissipation timescale $\tau_{\rm disk}$, we observe that satellites tend to survive more frequently when the dynamical tide is included. 
This effect is particularly pronounced at the highest satellite dissipation values considered. 
For $\Delta t_{\rm sat} = 1.0~\Delta t_\oplus$, no satellite survived in the equilibrium tide case (Fig.~\ref{Fig9}, first column).
However, in the dynamical tide case, the satellite initially located at $0.6~\aio$ survives at the largest final star-planet distances (second to fourth columns).
It is important to note that the maximum star-planet distance for the dynamical tide case exceeds that in the equilibrium tide case for the same disk lifetime.
Thus, disk lifetime alone is no longer a reliable proxy for planetary distance; instead, we must directly pay attention to the planet's final orbital distance.
As shown in Figure~\ref{Fig9}, one key reason why satellites survive even at the longest disk lifetime considered (50'000~yr) around an initially fast rotating star (second column), for satellite dissipation values $\leq 0.1~\Delta t_\oplus$, is the planet's final semi-major axis is significantly larger (0.1077~AU) than in the equilibrium tide case (0.035~AU, see first column). 

This behavior is illustrated in Figure~\ref{Fig10} for a disk lifetime of 35'000~yr, though the same trend holds for 50'000~yr. 
The planet's orbital evolution remains nearly identical between the equilibrium tide case (blue curve) and the dynamical case for the initially slow-rotating star (red curve). 
{Note that the equilibrium tide simulation corresponds to an initially fast rotating star, however due to the low dissipation of the equilibrium tide, the evolution of the planet is not as dependent on the initial spin as for the dynamical tide.}
{Conversely, for the dynamical tide in a slowly rotating star, the reduced dissipation is a consequence of the nature of the inertial waves it excites.}

{Indeed, a} slower stellar rotation leads to weaker dissipation, as demonstrated in previous studies \citep{2007ApJ...661.1180O,2016CeMDA.126..275B,2017A&A...604A.113B} {and as shown in Eq.~\ref{stellar_freq_av_diss} and \ref{epsilon_star} with the parameter $\hat{\epsilon}_\star^2$ which is proportional to the spin of the star squared $\Omega_\star^2$}.  

\begin{figure*}[!htbp]
\centerline{\includegraphics[width=0.9\linewidth]{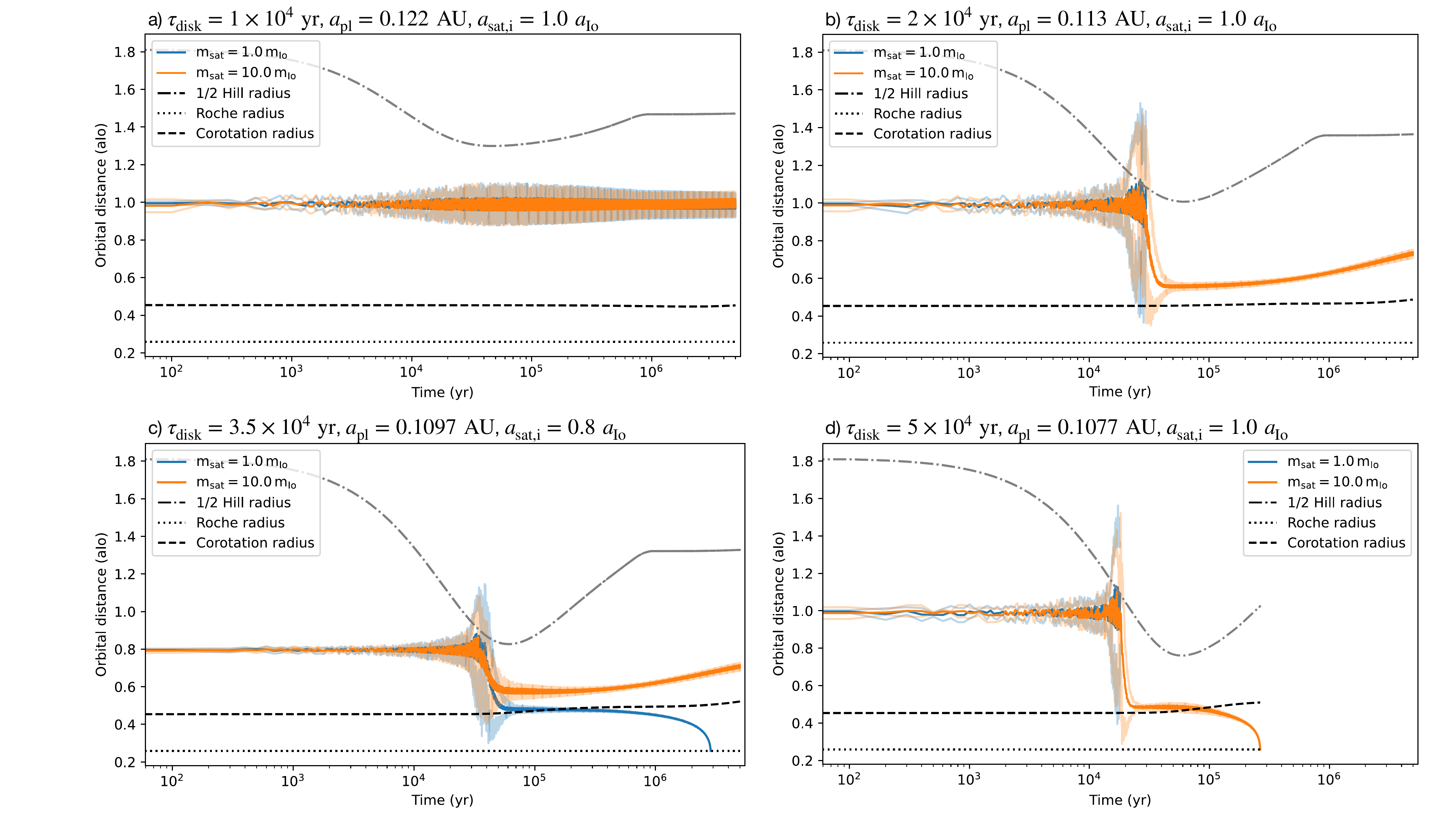}}
\caption{Evolution of the semi-major axis of a 1.0~$\Mio$ and a 10~$\Mio$ satellite on initially wide orbits for different disk lifetimes (i.e. different final planet distances). For all these cases, the initial rotation of the star is 1~day and the dissipation is $\Delta t_{\rm sat}=0.1~\Delta t_\oplus$ (which corresponds to the 2nd column, 4th row of Figs.~\ref{Fig9} and \ref{Fig12}). a) $\tau_{\rm disk} = 1\times 10^4$~yr, $a_{\rm sat,i} = 1.0~a_{\rm Io}$, b)  $\tau_{\rm disk} = 2\times 10^4$~yr, $a_{\rm sat,i} = 1.0~a_{\rm Io}$, c)  $\tau_{\rm disk} = 3.5\times 10^4$~yr, $a_{\rm sat,i} = 0.8~a_{\rm Io}$, d)  $\tau_{\rm disk} = 5\times 10^4$~yr, $a_{\rm sat,i} = 1.0~a_{\rm Io}$. These combinations were chosen to highlight some behaviors. }
\label{Fig11}
\end{figure*}

However, for an initially fast-rotating star, tides efficiently counteract disk-driven migration.
As shown in Fig.~\ref{Fig10} (left column), the planet initially migrates inwards, reaching $\sim$0.07~AU, before reversing direction and settling at its final position of $\sim${0.11~AU}.
This close approach to the star excites the eccentricity of the satellite, which in some cases is enough to induce dynamical disruption (particularly for low satellite dissipation values, see top left panel of Fig.~\ref{Fig10}, green curve). 
However, when dissipation is present, the eccentricity damping may allow the satellite to survive this planetary excursion. 
Several scenarios must be considered: 
\begin{itemize}
  \item If the satellite has no dissipation, eccentricity damping is instead driven by the planetary tide, which becomes effective as the planet moves away from the star, reducing the source of excitation (top left panel of Fig.~\ref{Fig10}, orange curve). Note that, since the dynamical tide is now included in the planet, planetary tide damping is significantly stronger than in the equilibrium tide case. This is one of the primary reasons why satellites survive more frequently in this scenario.
  \item If the satellite has some dissipation, it may experience tidal disruption as the planet approaches the star (bottom left panel of Fig.~\ref{Fig10}, {blue and red curves}). However, if the satellite remains outside corotation, it can survive, at least for the duration of the simulation (left panels of Fig.~\ref{Fig10}, {orange} curve).
\end{itemize}

Interestingly, while a stronger planetary tide enhances survival probabilities over the few Myr of the simulations, it is likely to prove detrimental in the long term.
As the planet's rotation evolves towards pseudo-synchronization, the corotation radius increases (dashed lines, left column of Fig.~\ref{Fig10}). 
For satellites orbiting an initially slow-rotating star, this transition occurs early (after $10^5$ years) causing the satellite to rapidly spiral inward, reaching the Roche limit in just over $3~\times 10^5~$yr.
Ultimately, all satellites will eventually cross the corotation radius, after which the strong planetary tide will drive inward migration, leading to their eventual collision with the planet.

For a given satellite dissipation, the boundary in final planet semi-major axis between survival and disruption (marked by the horizontal black dashed line in Fig.~\ref{Fig9}) remains consistent across the three rotation considered {(while being different in terms of disk lifetime)}.
{This means that the planet's final position is the dominant factor and, at the zeroth order, it does not matter how the planet gets there. 
A short disk lifetime and slow initial stellar rotation and a longer disk lifetime with a fast initial stellar rotation will lead to the same final planetary distance and the same outcome for the satellite (survival for instance) although their orbital evolution was different.
Indeed}, it appears to be of little consequence whether the planet reached its final position directly ($P_\star = 8$~day) or underwent a temporary inward excursion before settling ($P_\star = 1$~day and $P_\star = 3$~day). 
This is true, at least with our simulation's resolution. 
However, if we were to sample more finely in disk lifetime or final planet position, subtle differences might emerge.
In particular, the planet’s excursion closer to the star could play a role in satellite disruption, as observed in Figure~\ref{Fig10} (left column, red curves).

\section{What does this study teach us about the putative satellite around WASP-49 A b?}

Given the results from our tidal migration simulations, we briefly examine the claim of a volcanic satellite orbiting the hot Saturn WASP-49 A b ($\Mp = 0.365\pm0.019$~M$_{\mathrm{Jup}}$, $\Rp = 1.115\pm0.047~$R$_{\mathrm{Jup}}$) at a putative orbital period of $\sim$ 8 hours ($\sim$0.24~$\aio$), derived from variable alkali Doppler shift measurements \citep{2024ApJ...973L..53O}. 
The range of allowed satellite periods for this system is estimated to be 6-12 hours (corresponding to 0.19-0.31~$\aio$), based on the stability criterion of the Hill Radius ($\sim$0.49~R$_\mathrm{Hill}$ following \citealt{2006MNRAS.373.1227D}, and recently $\sim$ 0.41~R$_\mathrm{Hill}$ from \citealt{Kisare2024}).
This close-in orbit is compatible with a recent dynamical study considering planetary oblateness and stellar perturbations which showed a stability preference near (yet beyond) the Roche limit \citep{2025A&A...694L...8S}.
Satellite stability declines rapidly at larger satellite semi-major axes near the 1/2 Hill sphere criterion by \citet{2006MNRAS.373.1227D, Kisare2024}.
The hot Saturn itself orbits its Sun-like star ($0.894^{+0.039}_{-0.035}$~M$_\odot$) at 2.79 days, which corresponds to a semi-major axis of $\sim$ 0.039~AU.

This distance is similar to the shortest distance we consider in this study (0.035~AU), which is obtained when only considering the equilibrium tide in both star and planet (see Figures~\ref{Fig8} and \ref{Fig9}).
The possible case of the putative satellite of WASP-49 A b is therefore more similar to our ``equilibrium tide'' case, for the shortest star-planet distance (i.e. longer disk lifetime) and the shortest planet-satellite distance.
In these cases, regardless of satellite dissipation, we found that they do not survive the migration of the planet in the disk due to their proximity to the planet and the tides raised in the planet resulting in a tidal decay, which ultimately leads to engulfment.

Our system parameters differ slightly from those of WASP-49 A b. 
In particular, our Jupiter-like planet has a higher mass and a larger radius than the hot Saturn. 
{This is important for two main reasons: for interactions with the disk and for tidal interactions.
Indeed, a less massive planet ($0.365\pm0.019$~M$_{Jup}$) might not be able to open a gap in the circumstellar disk.
A quick calculation following the Eq. 38 of \citet{2004ApJ...604..388I} shows that a planet the mass of WASP-49 A b would open a gap in the disk for semi-major axes lower than 1.1~AU. 
Assuming the planet formed at distances higher than this value means that the circumplanetary disk was not isolated from the circumstellar disk and this could impact the formation of the satellite.
If the planet was formed at distance lower than 1.1~AU, then the satellite might have formed in an isolated circumplanetary disk before the planet reaches our initial simulation distance of 0.15~AU.
On the other hand, a higher planetary radius means a faster tidal evolution (all other parameters being equal).}
At an age of 1~Myr (our initial condition), our Jupiter has a radius of 1.48~R$_{\mathrm{Jup}}$ and at $\sim 6~$Myr (towards the end of the simulation), it remains around 1.41~R$_{\mathrm{Jup}}$ \citep{2013NatGe...6..347L}. 
{This value} is significantly larger than 1.115~R$_{\mathrm{Jup}}$, {the measured radius of WASP-49 A b} ({a value of 1.115~R$_{\mathrm{Jup}}$ corresponds to }the {modeled} radius of a Jupiter-like planet at an age of about 140~Myr, \citealt{2013NatGe...6..347L}). 
As a result, the planetary tide we obtain in our simulations is stronger than in the present-day WASP-49 A system. 

In any case, the formation of the system through type-II migration of the host planet appears highly challenging.
If the satellite's orbit was confirmed, specific simulations should be conducted using the system's parameters to explore why the planet may be less dissipative than expected (i.e., our assumed planetary $Q$ might be too low, leading to an overestimated dissipation rate).
In this work, we use a frequency-averaged tidal dissipation model (see Section~\ref{subsubsec:eq_dyn_tid_in_planet}). 
However, the actual dissipation of inertial waves exhibits a strong frequency dependence with peaks and troughs spanning several orders of magnitude \citep{2004ApJ...610..477O}.
If the excitation frequency induced by the satellite coincides with a dissipation trough, the evolutionary timescale becomes significantly longer, and the satellite's evolution would then be governed by the planet’s own evolution (as its radius contracts or as it spins up or down, shifting the locations of dissipation peaks and troughs, {\citealt{2021ApJ...918...16M}}).
This could be investigated with the latest version of \textsc{Posidonius} (\citealt{2024A&A...691L...3R}, Kwok et al. in prep, see next Section~\ref{sec:conclusion}).
Another possibility would be that the satellite was captured relatively ``recently'', though this scenario seems unlikely given the planet’s close proximity to its host star. 

Another scenario, recently discussed by \citet{2023MNRAS.520..761H}, involves the satellite becoming unbound. 
In our scenario, this would require both efficient outward migration and a highly dissipative satellite.
Outward migration would facilitate the moon’s escape, while high dissipation would regulate eccentricity growth, preventing excessive values that could otherwise trigger dynamical disruption.
In this case, the satellite could have formed, evolved, escaped the gravitational pull of the planet, only to later return and collide with the planet.
However, it seems unlikely we are observing the satellite in the final stages of this process, as the timescale for such a collision must be short compared to the system’s age. 

\section{Conclusions}\label{sec:conclusion}

In this study, we investigated the survival of satellites of masses $1~\Mio$ and $10~\Mio$ during the migration of their host planet to close-in distances, typical for hot Jupiters. 
Our primary conclusion is that satellite survival becomes highly unlikely when the planet migrates closer than $\approx 0.1$~AU. 
The key factors influencing satellite survival, in approximate order of importance, are: 1) the final star-planet distance; 2) the initial satellite-planet distance; 3) the satellite's tidal dissipation (where lower dissipation values enhances survival rate); 4) the mass of the satellite (with more massive satellites exhibiting slightly higher survivability).
Satellites that fail to survive are lost through two primary mechanisms: 1) {Dynamical disruption}, where eccentricity excitations drives the satellite's periastron below the Roche limit; 2) {Tidal disruption}, where orbital decay causes the satellite to gradually spiral inward until it reaches the Roche radius.
Even for satellites that do survive the few million years of evolution modeled in our simulations, long-term stability is not guaranteed. 
This uncertainty is particularly relevant since we did not consider tidal dissipation values as high as Io's (which is about $2.7\times \Delta t_\oplus$), suggesting that our estimates are rather conservative. 
In reality, satellite survivability may be even lower than what we predict here.

Ideally, studying the long-term survival of satellites would require significantly longer simulations, which are computationally expensive. 
As a result, our survival rates should be viewed as optimistic upper limits. 
As the system continues to evolve, the planet's rotation gradually shifts towards pseudo-synchronization (consistent with the CTL model we adopted).
The timescale for this evolution depends on the final star-planet distance, while the pseudo-synchronization period itself is determined by the planet's semi-major axis and eccentricity (which remains very small throughout our simulations).
Consequently, the corotation radius of the planet (which governs the direction of satellite migration) will continue to increase over time (as illustrated in Fig.~\ref{Fig5}, for instance).
This implies that all the satellites positioned above the corotation radius at the end of our simulations will eventually migrate below it as the system evolves further.
Once this transition occurs, both planetary tides and satellite tides (continuously driven by the eccentricity excitation caused by the nearby star), will work together to gradually decay the satellite's orbit.

In this study, we also explored the impact of the dynamical tide in both the star and the planet on 1) the evolution of the planet and 2) the evolution of the satellites it hosts. 
We identified several notable differences compared to cases where we considered only the equilibrium tide. 
In particular, massive satellites with high dissipation values ($1~\Delta t_\oplus$) failed to survive at star-planet distances lower than 0.125~AU under the equilibrium tide model. 
{And this should remain true for even higher satellite dissipation.}
However, when dynamical tides were included, these {dissipative }satellites were more likely to persist, with survival possible around planets as close as  0.117~AU (for an initially slow-rotating star).
This increased survival rate occurs because satellites that remain outside the corotation radius experience more efficient outward migration under the dynamical tide model, compared to the equilibrium tide case. 
{Given that the simulations including the dynamical tide are more realistic than the simulations with equilibrium tide only, this gives a small measure of hope for the survival of such satellites. 
Note however, than the parameter space of surviving satellites is very narrow (they all start their evolution at 0.6~$\aio$, not closer to the planet nor farther away).}

Furthermore, our results suggest that the proposed satellite around WASP-49 A b is unlikely to have survived planetary migration unless specific conditions were met. 
Given the planet's proximity to its host star and the typical tidal dissipation expected in gas giants, a satellite at the inferred orbital distance ($\sim$8 hours, $\sim$0.24~$\aio$) would likely have been lost through tidal decay. 
This raises the possibility that the satellite, if real, was captured relatively recently or that the planet and the satellite have a significantly lower dissipation than assumed in our models. 
Further targeted simulations using WASP-49 A b’s specific system parameters would be necessary to assess this scenario in more detail.

Beyond extending the duration of our simulations, this study could be further improved by incorporating a fully consistent treatment of tides, as was recently implemented in \citet{2024A&A...691L...3R}. 
\textsc{Posidonius} very recently included the capability to model both the proper tidal response of the equilibrium tide in rocky bodies \citep[allowed by][]{2024A&A...691L...3R} and the frequency-dependent dynamical tide response in gas giants and stars (Kwok et al. in prep). 
These new advancements would allow us to revisit the evolution of our systems, incorporating the full tidal spectrum of a gas giant planet \citep{2004ApJ...610..477O} or a star \citep{2007ApJ...661.1180O}, along with the more chaotic migration it would induce in both planet and satellite \citep[e.g.][]{2014A&A...561L...7A}.
{In addition, they would offer a more consistent treatment of the eccentricity than what we presented here (see discussion at the end of Section~\ref{subsec:eqdyntidestar}).}
This would allow us to explore the possibility that the satellite gets stuck in a trough of planetary dissipation, which then could help its survivability (at least with respect to the planetary dissipation, the satellite dissipation will always play a role due to the eccentricity excitation raised by the close-in star).
However, certain aspects of this improved tidal modeling (particularly for the dynamical tide in stars, Kwok et al. in prep) are still under validation/development.
Moreover, running simulations with these refinements enabled will require even greater computational resources and longer execution times than those required for this study. 
Thus, further investigation along these lines will be the focus of a future study.

{A fully consistent treatment of tides as described in the previous paragraph would also allow us to investigate the evolution of multiple satellites around the planet, for which a correct treatment of the eccentric tides is even more relevant. 
Indeed, just as Jupiter's inner satellites form a resonant chain (Mean Motion Resonances), exomoons could also form such chains in which the eccentricity is excited to higher levels than in a non-resonant system.
This eccentricity excitation would probably lead to a decrease in survivability of the satellites, especially that a new source of instabilities appear: satellite-satellite interactions and possible close-encounters.
If we were to study this question, we would probably have to consider planets stopping migration farther away from the star than in this study (i.e. where the parameter space in initial planet-satellite distance is wider, see Fig.~\ref{Fig9}).}

Additionally, we neglect here the tidal interaction between the star and the satellite. 
However, such an interaction might not be favorable for survival.
Indeed, synchronization timescales are very similar when we consider the tide raised by the planet compared to the tide raised by the star on the satellite (calculated taking the orbital distance of the planet).
This means that the two tides should be constantly fighting to synchronize the rotation of the satellite, which means that the satellite might never be synchronous with one or the other, which means then that there will always be some dissipation in the satellite.
This non-synchronization will enhance the inward migration due to the eccentricity of the satellite (see for instance Eq.3 of \citealt{2011A&A...535A..94B}).
To summarize, satellite survival during the inward migration of a hot Jupiter seems very challenging.

\section*{Data availability}

For this work, we used \textsc{Posidonius} version dbc5c8ec11e9055c0acfee91e012f39fe2fbc688 (\url{https://www.blancocuaresma.com/s/posidonius}). 
All initial conditions can be found on \url{https://zenodo.org/records/14930048} as well as the internal structure file that includes the inverse values of the structural quality factor ($1/Q'_{s,p}$ from Eq.~\ref{plan_freq_av_dissip}). 

\begin{acknowledgements}
The authors would like to thank the anonymous referee for their constructive comments.
This work was carried out within the framework of the NCCR PlanetS, supported by the Swiss National Science Foundation under grants 51NF40\_182901 and 51NF40\_205606.
E.B. acknowledges the financial support of the SNSF (grant number: 200021\_197176 and 200020\_215760).
C.M. acknowledges the support from the Swiss National Science Foundation under grant 200021\_204847 ``PlanetsInTime''.
The research described in this paper was carried out in part at the Jet Propulsion Laboratory, California Institute of Technology, under a contract with the National Aeronautics Space Administration. © 2025\\
The computations were performed at University of Geneva on the Baobab and Yggdrasil clusters.
This research has made use of the Astrophysics Data System, funded by NASA under Cooperative Agreement 80NSSC25M7105.
\end{acknowledgements}

\bibliographystyle{aa}

\begin{appendix}

\section{Validation of our implementation of the dynamical tide}\label{App_validation}

\begin{figure}[h]
\centerline{\includegraphics[width=\columnwidth]{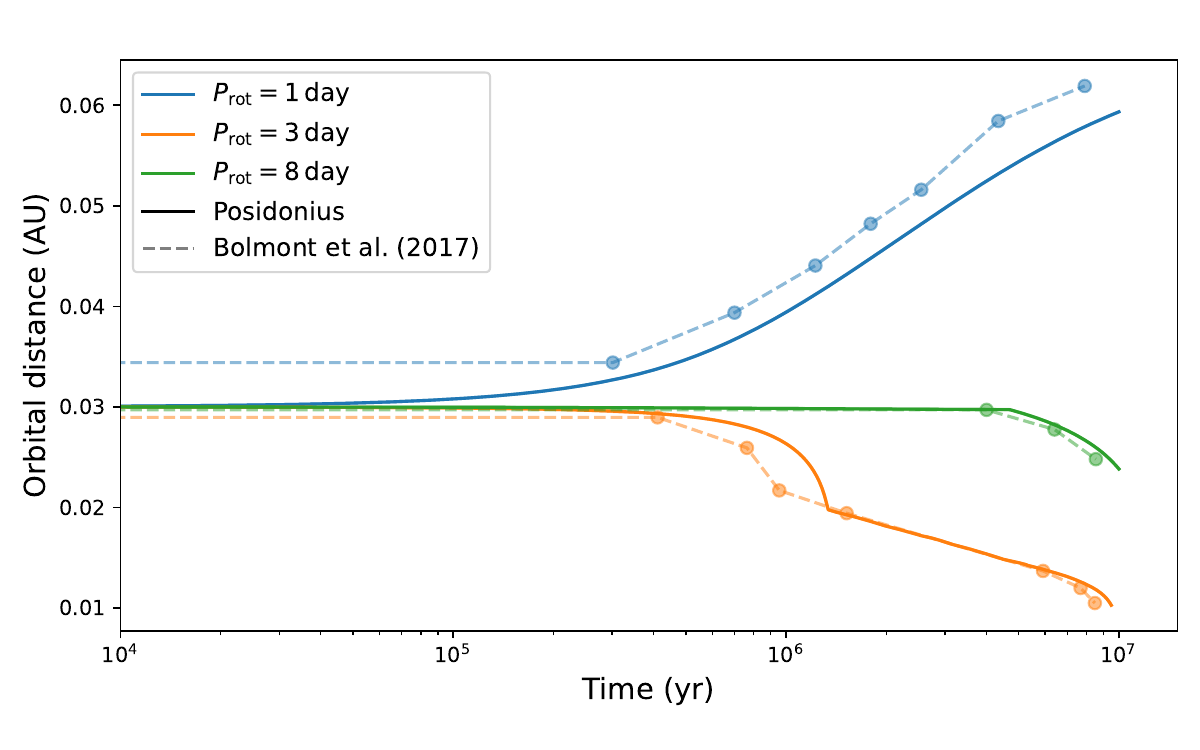}}
\caption{Comparison between the output of \textsc{Posidonius} and the Figure~6 of \citet{2017A&A...604A.113B} for the evolution of the semi-major axis of a Jupiter-mass planet due the dynamical tide and for different initial stellar rotation periods (1, 3 and 8~days). The dashed curves on this plot correspond the cyan curves of Figure~6 of \citet{2017A&A...604A.113B}, which were computed for a Sun metallicity (Z= 0.0134).}
\label{FigApp_validation}
\end{figure}

{Figure~\ref{FigApp_validation} shows a comparison of \textsc{Posidonius} simulations (ran for 10~Myr) compared to the Fig.6 of \citet{2017A&A...604A.113B}.
Our code reproduces well the behavior of \citet{2017A&A...604A.113B}, in particular, we capture the change from dynamical tide to equilibrium tide regime visible in the 3~day case (in orange) at around 1~Myr of evolution.
Due to the planet crossing the boundary corresponding to the frequency range of the excitation of the inertial waves ($n=2\Omega_\star$, where $n$ is the orbital frequency and $\Omega_\star$ the stellar spin), the dissipation decreases from high values to much lower values slowing down quite significantly the inward migration.
The planet then migrates inward following $n=2\Omega_\star$ until it is so close to the star that the equilibrium tide itself is enough to drive the inward migration. 
This behavior is explained in more details in \citet{2017A&A...604A.113B}.
The slightly different timescales are thought to come from some differences of numerical values for some constants in the different codes (Note that for \textsc{Posidonius}, the constants are referenced appropriately to ensure reproducibility whereas this was not done in \citet{2017A&A...604A.113B}).}

\section{A comparison between the various values of dissipation we consider and a tidal quality factor $Q$}\label{App1}

In a generic case (i.e. when multiple frequencies are excited in the system), there is no obvious correspondence between $\Delta t$ and a tidal quality factor $Q$ or $Q'=3/2\times Q/k_2$. 
The two models from which these parameters correspond (the Constant Time lag model for $\Delta t$ and the Constant Phase lag for $Q$) are fundamentally incompatible \citep[e.g.][]{2010A&A...516A..64L}.
However, a correspondence can be established if we restrict ourselves to a single excitation frequency. 
In the case of a purely circular coplanar orbit, this frequency is just $2|n-\Omega|$ (semi-diurnal frequency), where $n$ is the orbital frequency of the perturber and $\Omega$ the rotation frequency of the body in which the tides are operating.
If $n=\Omega$ (i.e., the deformed primary body is synchronously rotating), then the dominant tidal mode is the eccentric annual mode with a frequency of $n$.
Therefore, depending on whether we consider the planet or the satellite, the correspondence between a dissipation based on a $\Delta t$ or a $\sigma$ and a tidal quality factor $Q$ is different. 

\subsection{For the planet}

To compute a tidal quality factor $Q_\mathrm{p}'$ for the planet corresponding to the equilibrium tide, the expression is
\begin{align}\label{Eq:Qp_semi_diurnal}
    \frac{1}{Q_\mathrm{p}'} &= \frac{4}{3}k_\mathrm{2,p}\Delta t_p |n_{\mathrm{sat}}-\Omega_\mathrm{p}|,\\
    \quad & = 2 \sigma_\mathrm{p} \frac{\Rp^5}{\mathcal{G}} |n_{\mathrm{sat}}-\Omega_\mathrm{p}|,
\end{align}
where $n_{\mathrm{sat}}$ is the orbital frequency of the satellite and $\Omega_p$ is the spin of the planet. 
In our simulations, the initial rotation period of the planet is 13~hr.
We therefore calculate a value of $Q'_\mathrm{p}$ for the equilibrium tide of the planet with the value of $\sigma_p$ given in the main text Section~\ref{subsubsec:eq_dyn_tid_in_planet} for each initial satellite semi-major axis considered here.
The values are given in Table~\ref{App_tab:Q_p} for the initial age in our simulations (i.e. at an age of $10^6$~yr) and towards the end of it (i.e. at an age of $6\times10^6$~yr).

\begin{table}[h]
    \centering
    \caption{Values of the tidal quality factor $Q_\mathrm{p}$ for the planet for the equilibrium tide and dynamical tide.}
    \setlength{\tabcolsep}{3pt}
    \begin{tabular}{|c|c|c|c|c|c|c|}
        \cline{4-7}
        \multicolumn{1}{c}{}& \multicolumn{1}{c}{}& \multicolumn{1}{c|}{} & \multicolumn{2}{c|}{Equilibrium} & \multicolumn{2}{c|}{Dynamical} \\
        \multicolumn{1}{c}{}& \multicolumn{1}{c}{}& \multicolumn{1}{c|}{} & \multicolumn{2}{c|}{tide} & \multicolumn{2}{c|}{tide} \\
        \cline{1-7}

        \multicolumn{1}{|c|}{Age}& $\Rp$ & $k_{2,\mathrm{p}}$ & $Q_\mathrm{p}$ for & $Q_\mathrm{p}$ for & \multicolumn{2}{c|}{$Q_\mathrm{p}$} \\
        \multicolumn{1}{|c|}{(yr)}& ($\RJ$) &  & $0.3~\aio$ & $1.0~\aio$ & \multicolumn{2}{c|}{}\\
        \hline
        $10^6$ & 1.51 & 0.26 & $1.91\times10^5$ & $2.36\times10^5$ & \multicolumn{2}{c|}{$8.24\times10^4$}\\
        \hline
        $6\times10^6$ & 1.34 & 0.31 & $4.09\times10^5$ & $5.07\times10^5$ & \multicolumn{2}{c|}{$7.76\times10^4$}\\
        \hline
    \end{tabular}
    \tablefoot{Higher dissipation corresponds to lower values of $Q_p$. For all calculations here, we consider a rotation period of 13~hr (initial rotation period).}
    \label{App_tab:Q_p}
\end{table}

For the dynamical tide, we can directly compute the quality factor $Q'_\mathrm{p}$ from the structural quality factor $Q'_\mathrm{s,p}$ (of Equ.~\ref{plan_freq_av_dissip}) as follows
\begin{equation}
    \frac{1}{Q'_\mathrm{p}} = \frac{\hat{\epsilon_\mathrm{p}}^2}{Q'_\mathrm{s,p}},
\end{equation}
where $\hat{\epsilon_\mathrm{p}}^2 = \Omega_\mathrm{p}^2/\Omega_{\odot,c}$ as introduced in the main text (Sec.~\ref{subsubsec:eq_dyn_tid_in_planet}).
Using the initial rotation period of the planet which is 13~hr, we can compute a $Q'_\mathrm{p}$, and if we take the value of $k_2$ from the internal structure from \citet{2013NatGe...6..347L}, we can compute a $Q_\mathrm{p}$.
Table~\ref{App_tab:Q_p} also gives these values (last column). 
As expected, the tidal quality factor for the dynamical tide is lower than for the equilibrium tide, which leads to faster evolution. 
Our equilibrium tide tidal quality factor is however lower than the high values proposed for the equilibrium tide in the introduction ($Q_\mathrm{p}\sim 10^{10}-10^{12}$, \citealt{1977Icar...30..301G}). 
Considering a lower value would impact our simulations for a non-dissipative satellite, however for realistic values of the satellite dissipation ($\Delta t_\mathrm{sat}\sim 1~\Delta t_\oplus$) the satellites' fate is mainly due to satellite tide.

\subsection{For the satellite}

For the satellite, as its rotation very quickly evolves towards tidal locking ($n_{\mathrm{sat}}=\Omega_\mathrm{sat}$), we cannot use Equation~\ref{Eq:Qp_semi_diurnal}. 
In that case, the main tidal mode is the eccentric annual mode of frequency $n$ so that 
\begin{align}\label{Eq:Qp_diurnal}
    \frac{1}{Q_\mathrm{sat}'} &= \frac{2}{3}k_{2,\mathrm{sat}}\Delta t_\mathrm{sat} n = \sigma_\mathrm{sat} \frac{R_\mathrm{sat}^5}{\mathcal{G}} n.
\end{align}
The values for the dissipation in the $1~\Mio$ satellite are given in Table~\ref{App_tab:Q_sat}.
The values for $\Delta t_\mathrm{sat} = 1.0~\Delta t_\oplus$ are compatible with the recent values of \citet{Park2025} of $Q_\mathrm{Io} = 11.4\pm 3.6$.

\begin{table}[h]
    \centering
    \caption{Values of the tidal quality factor $Q_\mathrm{sat}$ for the satellites.}
    \begin{tabular}{|c|c|c|}
        \cline{2-3}
        \multicolumn{1}{c|}{}& $Q_\mathrm{sat}$ for & $Q_\mathrm{sat}$ for  \\
        \multicolumn{1}{c|}{}& $0.3~\aio$ & $1.0~\aio$ \\
        \hline
        $\Delta t_\mathrm{sat} = 10^{-4}~\Delta t_\oplus$ & $6.28\times10^4$ & $3.82\times10^5$  \\
        \hline
        $\Delta t_\mathrm{sat} = 1.0~\Delta t_\oplus$ & 6.28 & 38.2  \\
        \hline
    \end{tabular}
    \tablefoot{We give here the lowest and the highest values of dissipation we consider. Higher dissipation corresponds to lower values of $Q_\mathrm{sat}$.}
    \label{App_tab:Q_sat}
\end{table}

\section{Evolution of the planet's rotation}

{Figure~\ref{FigApp_rotation_planet} shows the evolution of the spin of the planet towards pseudo-synchronization for two different cases. 

\begin{figure}[h]
\centerline{\includegraphics[width=0.8\columnwidth]{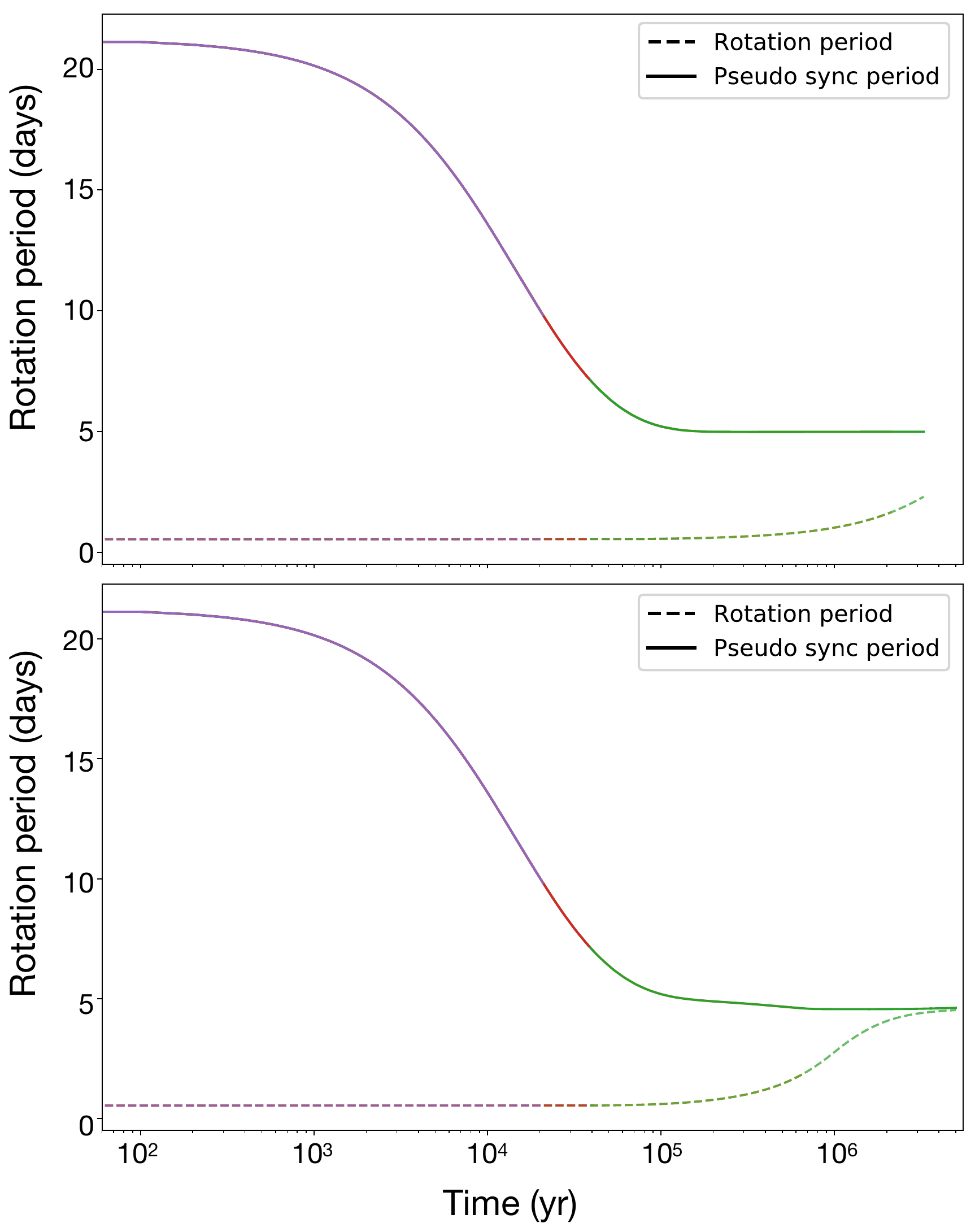}}
\caption{{Rotation period of the planet (dashed lines) and pseudo-synchronization period (full lines) as a function of time. The top panel corresponds to the simulation of Figure~\ref{Fig5}: $\tau_{\rm disk} = 3\times10^4$~yr, 1~$\Mio$, equilibrium tide only. The bottom panel corresponds to: $\tau_{\rm disk} = 3\times10^4$~yr, 1~$\Mio$, initial stellar rotation of 8~day and the dynamical tide is on in both star and planet. 
The different colors correspond to the different satellite initial semi-major axis. They are all superimposed and the green one corresponds at an initial satellite semi-major axis of $0.6~\aio$.}}
\label{FigApp_rotation_planet}
\end{figure}

The first one (top panel) corresponds to the simulation of Figure~\ref{Fig5}: $\tau_{\rm disk} = 3\times10^4$~yr, 1~$\Mio$, equilibrium tide only, while the second one (bottom panel) corresponds to: $\tau_{\rm disk} = 3\times10^4$~yr, 1~$\Mio$, dynamical tide in both star and planet and an initial stellar rotation of 8~day.
In the first configuration, the planet stops its disk-induced migration at about 0.06~AU, while it stops at about 0.08~AU for the second.} 

{In both cases, we see the pseudo-synchronization period (which is a function of the orbital frequency $n$ of the planet and its eccentricity, here close to zero) sharply decreasing due to the disk-induced inward migration. 
It reaches a final value of about 5 days once the migration has stopped.
The planet starts with a fast rotation (13 hours) and gradually slows down towards the pseudo-synchronization period.}

{In the top panel, the rotation of the planet does not reach the pseudo-synchronization period before the end of our simulation.
In the bottom panel, the rotation of the planet evolves faster due to the enhanced dissipation in the planet due to the dynamical tide, and reaches pseudo-synchronization within 6~Myr.}

\section{Impact of satellite mass for dynamical tide simulations}\label{App_10Mio}

\begin{figure*}[!htbp]
\centerline{\includegraphics[width=0.9\linewidth]{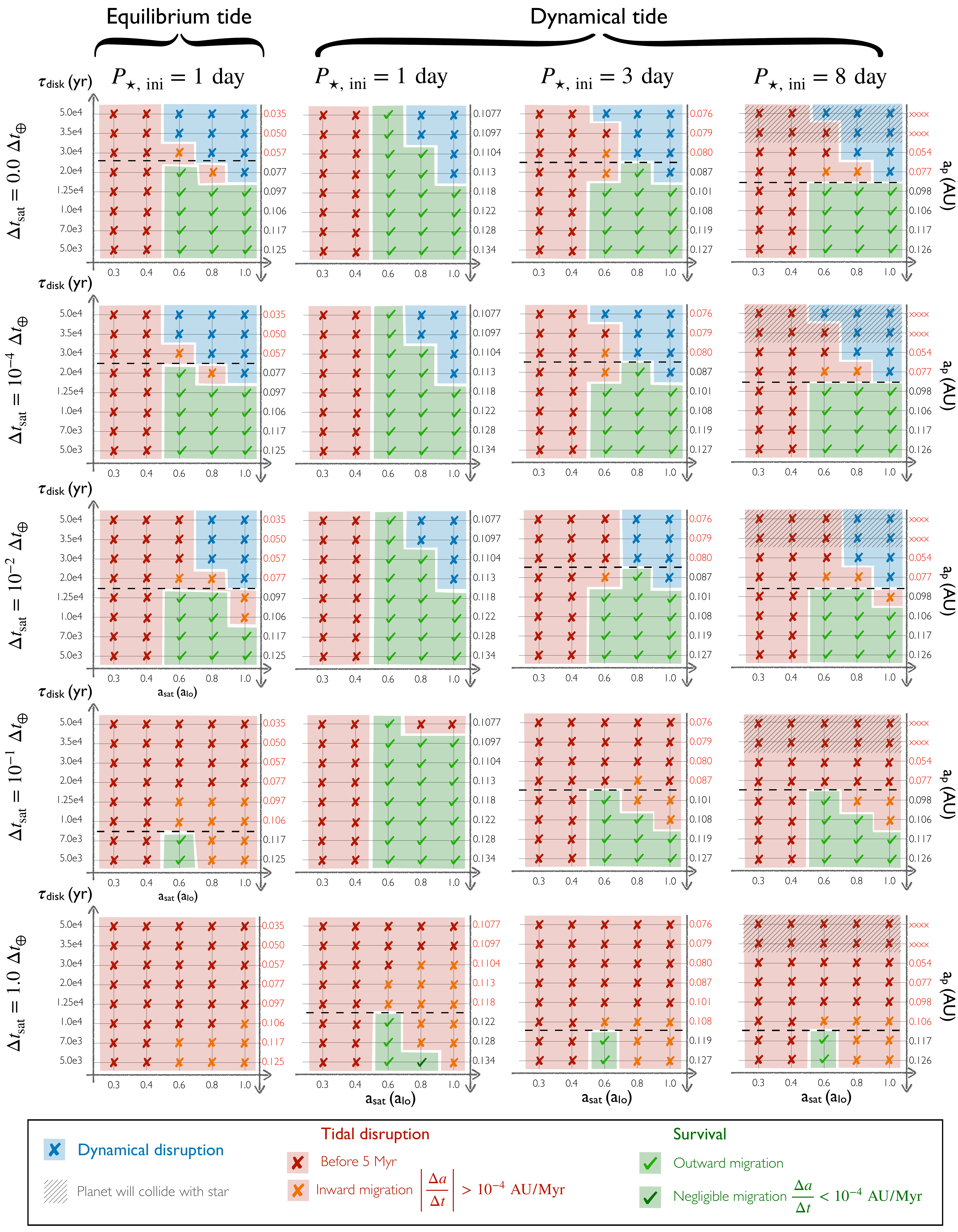}}
\caption{Same as Figure~\ref{Fig9} but for a satellite of 10~$\Mio$. }
\label{Fig12}
\end{figure*}

As discussed in Section~\ref{Results_Eq_mass_sat}, increasing the mass of the satellite reduces its susceptibility to dynamical disruption. 
This trend is evident when comparing Figures~\ref{Fig9} and \ref{Fig12}, where the blue region (representing dynamically disrupted satellites) shrinks as satellite mass increases from $1~\Mio$ to $10~\Mio$. 
For 1~$\Mio$, dynamical disruption was still possible at $0.1\times \Delta t_\oplus$, whereas for 10~$\Mio$, only tidal disruption occurs. 
This distinction is further illustrated in Figure~\ref{Fig11}, which presents examples of the orbital evolution for 1~$\Mio$ and 10~$\Mio$ satellites at $\Delta t_{\rm sat} = 0.1\times \Delta t_\oplus$, with an initial stellar rotation period of 1~day and varying disk lifetimes (corresponding to the second column and fourth row of Figs.~\ref{Fig9} and \ref{Fig12}). 
In particular, Figure~\ref{Fig11}d (bottom right panel), shows the evolution for the longest disk lifetime and a satellite initially located at $1~\aio$. 
While the 1~$\Mio$ satellite undergoes dynamical disruption early in the evolution, {the} 10~$\Mio$ satellite experiences rapid eccentricity damping due to enhanced dissipation, leading to inward migration. 
Once the eccentricity is sufficiently damped, the inward migration slows down but then accelerates again when the satellite crosses the corotation radius, eventually reaching the Roche limit after a few $10^5$~yrs.
This kind of evolution, predominantly governed by tides in the satellite, closely resembles the behavior observed in the equilibrium-tide only scenario (see Section~\ref{Results_Eq_mass_sat}).

Additionally, more massive satellites are less prone to tidal disruption, allowing them to survive around planets on closer orbits compared to their lower-mass counterparts.
This effect is particularly evident for a dissipation of 0.1~$\Delta t_\oplus$ and for a satellite initially positioned farthest from the planet (1~$\aio$). 
This scenario still corresponds to the second column, fourth row of Figs.~\ref{Fig9} and \ref{Fig12}.
By comparing these two figures, we observe that $1~\Mio$ satellites can only survive if the planet remains at a distance of $\ge$0.128~AU from the star.
However, for $10~\Mio$ satellites, survival is possible at slightly smaller separations, with the planet at $\ge$0.1097~AU from the star.
 
An example of orbital evolution for a disk lifetime of $1\times10^4$ is shown in Figure~\ref{Fig11}, which results in tidal disruption for the $1~\Mio$ satellite (slight inward decay, orange cross in Fig.~\ref{Fig9}), but survival for the $10~\Mio$ satellite (slight outward migration, light green cross in Fig.~\ref{Fig12}).
Similarly, Figure~\ref{Fig11}b illustrates an evolution for a disk lifetime of $2\times10^4$, leading to dynamical disruption for the $1~\Mio$ satellite (blue cross in Fig.~\ref{Fig9}), but survival for the $10~\Mio$ satellite (light green cross in Fig.~\ref{Fig12}).
In summary, Figure~\ref{Fig11} shows that $1~\Mio$ satellites are either:
\begin{itemize}
  \item Dynamically disrupted for disk lifetimes $\ge2\times10^4$ yr (corresponding to star-planet distances $\le 0.113$~AU), or 
  \item Experience eccentricity excitation that gradually drives them towards the planet for disk lifetimes between $10^4$ and $1.25\times10^4$ yr (corresponding to star-planet distances $0.118$ and $0.122$~AU).
\end{itemize}

As discussed earlier, $10~\Mio$ satellites can escape dynamical disruption for $\tau_\mathrm{disk}$ = 2, 3, 3.5$\times10^4$ yr by rapidly reducing their eccentricity due to the strong tide raised within them.
Once their eccentricity has sufficiently decreased, they undergo strong planetary-tide induced outward migration, which allows them to survive. 
This strong outward migration also prevents their orbit from slowly decaying, particularly for disk lifetimes between $10^4$ and $1.25\times10^4$ yr. 

As with the $1~\Mio$ satellite, the initial stellar rotation has a minimal impact on survival. 
While variations exist for a given disk lifetime, a more consistent pattern emerges when focusing instead on the planet’s final distance from the star.
Although more massive satellites survive around planets which are closer to their star than their $1~\Mio$ counterparts, the overall limit remains of the same order of magnitude between the two masses. 
Satellites closer than 0.1~AU are unlikely to survive.
The increased survivability of massive satellites is primarily due to efficient outward migration induced by the planetary tide (as seen in Fig.~\ref{Fig11}b and c) when they remain outside corotation. 
However, this also means that their eccentricity becomes more excited as they approach the Hill radius, eventually leading to a slow tidal decay when: 1) the planetary tide weakens as the planet’s radius decreases over time, 2) the corotation radius increases sufficiently to reverse the planetary tide-driven migration. 
Thus, our survival rates should be regarded as optimistic upper limits.

For the $10~\Mio$ satellites, the effect of the initial stellar rotation is slightly more pronounced compared to $1~\Mio$ satellites. 
This distinction is particularly {noticeable} at the high satellite dissipation (last row of Fig.~\ref{Fig12}).
This indicates that for initially slow-rotating stars, the survival limit can shift slightly inward.
Specifically, for an initial stellar rotation period of 1~day, while the satellite survives only if the planet remains beyond 0.122~AU, it does not persist at 0.118~AU. 
However, for a slower initial stellar rotation period of 8 days, the satellite survives at slightly smaller orbital separations, remaining stable as long as the planet is beyond 0.117~AU. 
Nevertheless, the overall effect remains small.

\end{appendix}
\end{document}